\documentclass[twocolumn]{aastex62}
\usepackage{rotating,multirow}
\usepackage{color}

\graphicspath{{./}{figures/}}
\newcommand\arcdegr{\mbox{$^\circ$}} 

\accepted{June 22, 2019}

\shorttitle{Constraints from Neutrino Emission of TXS~0506+056}
\shortauthors{Reimer et al.}

\begin{document}

\title{Cascading Constraints from Neutrino Emitting Blazars: The case of TXS~0506+056}

\email{Anita.Reimer@uibk.ac.at, Markus.Bottcher@nwu.ac.za,} 
\email{buson@astro.uni-wuerzburg.de}

\author{Anita Reimer}
\affiliation{Institute for Astro- \& Particle Physics, 
University of Innsbruck, 6020 Innsbruck, Austria}

\author{Markus B\"ottcher}
\affiliation{Centre for Space Research, North-West University, Potchefstroom, 2531, South Africa}

\author{Sara Buson}
\affiliation{NASA Postdoctoral Program Fellow, Universities Space Research Association, USA
\\
based at NASA Goddard Space Flight Center, Greenbelt, MD 20771, USA
\\
now at University of W\"{u}rzburg, 97074 W\"{u}rzburg, Germany}




\begin{abstract}
We present a procedure to generally constrain the environments of neutrino-producing
sites in photomeson production models of jetted Active Galactic Nuclei (AGN) where any origin of the dominant target photon field can be accommodated.
For this purpose we reconstruct the minimum target photon spectrum required to produce the (observed) neutrino spectrum, and
derive the distributions of all corresponding secondary particles. These initiate electromagnetic cascades with an efficiency that is linked to the neutrino production rate. The derived photon spectra represent the minimum radiation emerging from
the source that is strictly associated with the photo-hadronically produced neutrinos.

Using the 2014--15 neutrino spectrum observed by IceCube from TXS~0506+056,
we conduct a comprehensive study of these cascade spectra and compare them to the simultaneous multi-wavelength emission.
For this set of observations, photopion production from a co-spatially produced (co-moving) photon target can be ruled out as well as a setup
where synchrotron or Compton-synchrotron supported cascades on a stationary (AGN rest frame) target photon field operate in this source.
However, a scenario where Compton-driven cascades develop in the stationary soft-X-ray photon target which photo-hadronically
produced the observed neutrinos appears feasible with required proton kinetic
jet powers near the Eddington limit. The source is then found to produce neutrinos inefficiently,
and emits GeV photons significantly below the observed {\it Fermi}-LAT-flux. Hence, the neutrinos and the bulk of the gamma rays observed in 2014/2015 from TXS~0506+056 cannot have been initiated by the same process.

\end{abstract}


\section{Introduction}
\label{sec:intro}
Active Galactic Nuclei (AGN) have long been considered to be among the source populations responsible for 
the ultra-high energy cosmic rays observed at Earth. Indeed, members of this source population fulfill the 
Hillas criterion \citep[][]{Hillas84} as well as energetics requirements to accelerate 
charged particles to such extreme energies, in principle. These relativistic hadrons interact in the highly 
radiative environment of AGN, if above the threshold for photomeson and/or Bethe-Heitler pair production, 
or interact in dense material associated with the AGN via inelastic nucleon-nucleon interactions, 
to produce 
secondary particles, among them electron-positron pairs, neutrinos and high-energy photons. 
Hence, for just as long, gamma-ray loud AGN have been predicted as sources of neutrinos 
\citep{Mannheim89,Stecker91,Mannheim92,Protheroe92,Mannheim93,Szabo94,Mastichiadis96,Bednarek97,Bednarek99,
Protheroe99,Muecke01,Mannheim01,Atoyan01,Protheroe03,Atoyan03,Muecke03,Reimer04,Dermer09,Dimi12,Dermer12,
Boettcher13,Halzen13,Murase14,Dermer14,Gao17,Murase18}.

Until a few years ago, multi-messenger astrophysics was merely a prediction by (some) cosmic-ray 
theorists. It turned into reality with the first detections of high-energy neutrinos of astrophysical 
origin \citep{IceCubeDiscovery2013}. Observationally, no individual sources could be associated with 
these  
neutrino events, which were found to be isotropically 
distributed over the sky \citep{IceCubeIsotropy2014}. Most interestingly, the total diffuse flux of 
these neutrinos follows a differential spectrum not harder than $dN/dE \propto E^{-2}$ above a few 
tens of TeV and ranging to almost 10 PeV \citep{IceCubeAllsky2015}. The detection of these neutrinos 
initiated an avalanche of hadronic emission models for various possible high-energy source populations, 
among them blazars, a sub-class of jetted AGN with the line-of-sight close to the jet axis. Blazars are 
the most numerous sources in the extragalactic GeV-TeV $\gamma$-ray sky \citep[e.g.,][]{ReimerReview,3FGL,2FHL}. 
The fact that in hadronic interactions, roughly equal powers of high-energy photons + pairs and 
neutrinos are produced \citep[e.g.,][]{MannheimSchlickeiser94,SOPHIA} makes these $\gamma$-ray 
blazars plausible sources of the detected neutrinos 
\citep[e.g.,][]{Dimitrakoudis14,Krauss2014,Cerruti15,Padovani15,Petropoulou15,Kadler2016}. 
One-zone blazar emission models, where neutrinos are produced in photo-hadronic interactions, 
typically predict neutrino spectra peaking at or beyond PeV-energies, with extremely hard spectral 
shapes below \citep[e.g.,][]{Muecke03,Dermer12} when applied to GeV $\gamma$-ray sources.

Recently, neutrino astrophysics experienced a further boost when the BL~Lac object TXS~0506+056 
was found within the 50~\% containment region of the single IceCube event IceCube-EHE-170922A 
\citep{IceCube1}, and simultaneously in an elevated flux state in the {\it Fermi}-LAT energy range 
\citep{IceCube1} at the time this event was detected. Furthermore, an extended neutrino excess 
(designated the ``neutrino flare'' in the following) consisting of $\sim 13$ neutrino events detected in 
a $110\pm^{35}_{24}$ days period in September 2014 -- March 2015 was found from the same source 
\citep{IceCube2}. This has motivated the suggestion that TXS~0506+056 is the source of these 
detected neutrinos \citep{IceCube1}, implicitly assuming a causal connection between the 
neutrino and $\gamma$-ray emission. Such a link is justified if at least part of these 
$\geq$~GeV-photons have been initiated by the same process responsible for the neutrino 
flux, or have at least been produced in the same emission region of the 
source\footnote{In this case co-acceleration of electrons and protons
can be expected to provide a causal connection between photon and neutrino production.}. 
It is, however, questionable whether a $\geq$GeV $\gamma$-ray emitting source can 
simultaneously be an efficient producer of neutrinos in the 10 -- 100~TeV energy 
range, in the framework of a photo-hadronic emission model \citep[e.g.,][]{Begelman90}. 
The latter requires a photomeson production optical depth $\tau_{p\gamma}\gg 1$ in a 
field of target photons at UV-to-X-ray energies (see Sect.~\ref{sec:NeutrinoProduction}),
while the former requires GeV-photons to escape the source, hence the $\gamma\gamma$-pair 
production optical depth $\tau_{\gamma\gamma} \ll 1$ for target photons at X-ray energies. Hence, 
the same target radiation field necessary for neutrino production is also a source of opacity 
for $\geq$GeV photons. Observe that a ratio of  $\tau_{p\gamma}/\tau_{\gamma\gamma} \gg 1$ 
is required in order for the radiation to escape, while $\tau_{p\gamma}/\tau_{\gamma\gamma}\approx 
\sigma_{p\gamma}/\sigma_{\gamma\gamma} \ll 1$ is guaranteed by the nature of the processes.

In this work we present a procedure that provides, in the framework of photo-hadronically 
produced neutrinos in jetted AGN, constraints on their associated broadband photon spectral 
energy distribution (SED) in a way that is independent of the origin of the dominant target 
photon field. It is expected to guide searches for neutrino source associations with electromagnetic 
counterparts for cases of highly radiative sources where neutrinos are produced predominantly 
photo-hadronically. We present this procedure by applying it to the 2014 -- 15 neutrino flare of 
TXS~0506+056. The resulting constraints are then discussed to infer limits on the required source 
power and multi-wavelength (MWL) predictions of the photon SED. The comparison to MWL data allows us in turn to 
constrain models.

In the following, we first (Sect.~\ref{sec:MWLdata}) present the quasi-simultaneous multi-messenger 
data acquired for the time period of the 2014 -- 15 neutrino flare, to be placed into a generic 
theoretical set-up where neutrinos are produced photo-hadronically in a relativistic jet. The 
subsequent Section \ref{sec:Method} describes the simulations of electromagnetic cascades initiated
by photo-hadronic processes and the constraints that can be deduced from them. Section \ref{sec:Implications}
then discusses implications for the nature of the target photon field for photo-hadronic processes
responsible for the neutrino emission, where we will show that the only plausible scenario
of photo-pion production in the jet of TXS~0506+056 requires a stationary UV -- X-ray target photon
field external to the jet that leads to Compton-supported photo-pion-induced cascades. We find that in this set-up the
neutrino-production optical depth is surprisingly low during the neutrino flare unlike the case of efficient neutrino production, $\tau_{p\gamma}\gg 1$. This in turn allows photon escape of any origin from the neutrino 
production site only below $\sim 10^{-5} E_{\nu,\rm obs}$ (with $E_{\nu,\rm obs}$ the observed neutrino energy). 
The $\gamma$-ray flux strictly associated with the neutrino flare is shown to lie significantly below the
GeV-flux detected contemporaneous with the IceCube neutrino
flare, implying no common production origin between the observed IceCube neutrinos and LAT gamma rays.
In Section \ref{sec:summary} we provide a summary of our results and conclusions. Throughout the paper, 
primed quantities refer to the 
co-moving jet frame of the emission region. Physical quantities are parameterized as 
$Q = 10^i \, Q_i $ in c.g.s units.

\section{Quasi-simultaneuous multi-messenger data of the jet}
\label{sec:MWLdata}

\subsection{Neutrino data}
\label{sec:NeutrinoData}

The neutrino flux from the 2014 -- 15 neutrino flare of TXS~0506+056 was best described by a 
spectrum between 32~TeV and 3.6~PeV of the form $\Phi_{\nu} (E_{\nu}) = \Phi_0 \, E_{14}^{-2.1}$ 
with $\Phi_0 = 2.2 \times 10^{-15}$~cm$^{-2}$~s$^{-1}$~TeV$^{-1}$ and $E_{14} \equiv E_{\nu} / 
(100 \, {\rm TeV})$ \citep{IceCube2}. For a redshift of TXS~0506+056 as measured by \citet{Ajello14}
and \citet{Paiano18}, 
$z = 0.3365$, which corresponds to a luminosity distance of $d_L \approx 1.8$~Gpc~$\approx 5.5 
\times 10^{27}$~cm, the IceCube neutrino flux corresponds to an isotropic-equivalent luminosity 
in muon neutrinos of $L_{\nu, \rm iso} \approx 5.8 \times 10^{46}$~erg~s$^{-1}$.

We assume that the neutrinos are produced by relativistic hadrons in the high-energy emission region 
(sometimes referred to as the 'blob') of (jet-frame) size $R'$, moving relativistically 
along the jet with constant Lorentz factor $\Gamma \equiv 10 \, \Gamma_{10}$ and viewing angle 
$\theta_{\rm obs}$, resulting in relativistic Doppler boosting by the Doppler factor $D = 
\left( \Gamma \, \left[ 1 - \beta_{\Gamma} \cos\theta_{\rm obs} \right] \right)^{-1} \equiv 10 \, D_{1}$ with
 $\beta_{\Gamma}c$ the bulk velocity.
In that case, the total neutrino luminosity (comprising all neutrino flavors and assuming 
complete flavor mixing) produced in the co-moving frame of the emission region is reduced to 
$L'_{\nu} \approx 1.7 \times 10^{43} \, D_{1}^{-4}$~erg~s$^{-1}$.

\subsection{Photon data}
To study the electromagnetic behaviour of TXS~0506+056 during the neutrino flare we collect 
 data in the optical, X-rays and $\gamma$-ray bands. 
Optical {\it V}-band and {\it g}-band data are obtained from the All-Sky Automated Survey for Supernovae 
\citep{ASAS-SN1,ASAS-SN2}. Data are publicly available from the ASAS-SN website\footnote{https://asas-sn.osu.edu/} 
and available between 2014 February 2 -- 2018 September 25.
At X-ray energies, we use data from the Burst Alert Telescope (BAT), on board
the Neil Gehrels {\it Swift} Observatory, to derive a constraint
at keV energies for a time interval simultaneous with the neutrino flare.
The $Fermi$-Large Area Telescope (LAT) has been monitoring the whole sky for almost ten years, including the 
relevant period of the neutrino flare as described in \ref{sec:LATdata}.

\subsubsection{{\it Swift-BAT} data analysis}
\label{sec:SWIFTdata}
The {\it BAT} survey data  were retrieved from the HEASARC public
archive\footnote{http://heasarc.gsfc.nasa.gov/docs/archive.html}
and  processed using the {\sc batimager} code \citep{segreto10},
dedicated to the processing of coded mask instrument data, that performs
processing and cleaning of the data, source detection,
and production of background subtracted images and other
scientific products.
TXS~0506+056 is not detected in the whole set of survey data, spanning from December 2004 to November 2017. Using 
the $20-85$~keV sensitivity map, for the time period spanning September 16, 2014 to February 28, 2017 we have 
derived an upper limit of $9.12\cdot 10^{-12}$~erg~cm$^{-2}$~s$^{-1}$
to its flux, at a $5\sigma$ level.

\subsubsection{Fermi-LAT data}
\label{sec:LATdata}

In this work we consider ten years of {\it Fermi}-LAT data, collected between 2008 August 4 and 2018 September 29 
(MJD 54682-58390). Events are selected in a $10\arcdegr 
\times 10\arcdegr$ square centered on the position of TXS~0506+056, from 100~MeV up to 300~GeV. To minimize the 
contamination from $\gamma$ rays produced in the Earth's atmosphere, we apply a zenith angle cut of  
$\theta<90\arcdegr$. 
Time periods during which the LAT has detected bright solar flares and $\gamma$-ray bursts were excluded. 
We perform a 
binned likelihood analysis with the package fermipy\footnote{\url{https://fermipy.readthedocs.io}} (v0.17.3),
based on the standard LAT 
ScienceTools\footnote{\url{http://fermi.gsfc.nasa.gov/ssc/data/analysis/}} (v11-07-00) and the P8R2\_SOURCE\_V6 
instrument response functions. The model of the region accounts for all sources included in the \emph{Fermi}-LAT 
preliminary 8-year source list\footnote{FL8Y preliminary catalog: 
\url{https://fermi.gsfc.nasa.gov/ssc/data/access/lat/fl8y/}}, 
and contained in a 15$\arcdegr$ region from the position of TXS~0506+056, as well as the isotropic and 
Galactic diffuse 
$\gamma$-ray emission models (iso\_P8R2\_SOURCE V6 v06.txt and gll\_iem\_v06.fits). The spectral 
parameters of the bright 
$\gamma$-ray object located $1.2\arcdegr$ from TXS~0506+056 and associated with the blazar 
PKS 0502+049, are left free 
to vary in the fit. Similarly to \citet{Garrappa_inprep}, the residual map of the region 
does not show evidence for 
significant structures, indicating that our best-fit model adequately describes the region.

To investigate the $\gamma$-ray data of TXS~0506+056 during the neutrino flare, we adopt 
as definition for the time 
interval the ``box window'' identified by \citet{IceCube2}, i.e. 158 days (MJD 56937 
to MJD 57096, that is 2014-10-07 to 2015-03-15). For this period, 
the $\gamma$-ray SED of TXS~0506+056 is well modelled with a power-law spectral shape. 
A maximum likelihood fit yields 
the best-fit ($>$ 100\,MeV) flux value of $(3.3 \pm 0.9)\cdot 10^{-8}$cm$^{-2}$s$^{-1}$, 
with a photon spectral index 
of $1.9 \pm 0.1$, and the source is detected at a significance of more than 12$\sigma$.
This gives an isotropic equivalent gamma-ray luminosity in the LAT-energy range during 
the neutrino flare 
of $1.9\cdot 10^{45}$erg~s$^{-1}$, or $1.9\cdot 10^{41} D_1^{-4}$erg s$^{-1}$ in the co-moving 
frame of the jet.

Downturns or upturns in the MeV-GeV spectrum of  TXS~0506+056 could be interesting to 
confirm / rule out features generated by the cascade scenarios. A previous work reported a hint 
for a hardening of the TXS~0506+056 spectrum in the $>2$ GeV LAT band \citep{Padovani18} during 
the 2014-15 "neutrino flare" interval. Issues arising from source confusion at the lower LAT 
energies prevented the same authors from conducting a full investigation of the blazar spectrum 
in the $>$100 MeV range.
In our analysis we are able to overcome these issues by accounting for the nearby known bright 
gamma-ray blazar  PKS~0502+049 in our model for the region of interest, and allowing its spectral 
parameters to vary in the fit. During the neutrino flare, we find that the source flux is consistent 
with a quiescent state and the spectral shape with the average one in the $>100$ MeV range 
\citep[see also][]{Garrappa_inprep}. Including this full broad-band information provided by 
the LAT allows us to rule out a large variety of cascade scenarios, as explained in more detail 
in Sec. \ref{sec:Cascading}.

\begin{figure}
\centering
  \includegraphics[width=0.48\textwidth]{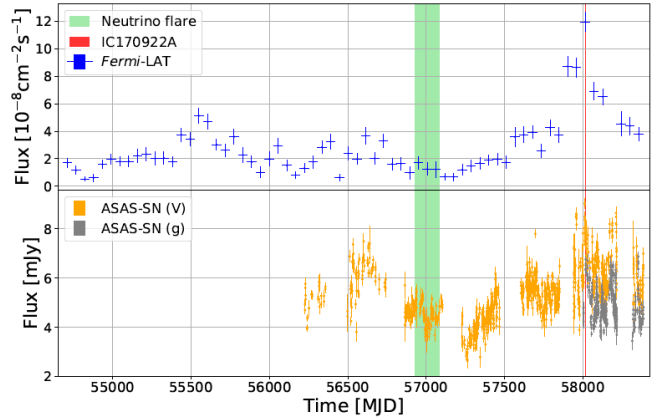}
\caption{Top: \emph{Fermi}-LAT light curve (photon flux integrated above 300~MeV) of 
TXS~0506+056 since mission start 
using 8-weeks binning. Bottom: Optical light curve from ASAS-SN in the V (yellow) 
and $\gamma$-ray (gray) bands. The 
green shaded area indicates the time interval of the neutrino flare while the red 
line corresponds to the detection 
time of the IceCube-EHE-170922A event.}
\label{fig:0}   
\end{figure}

Fig. \ref{fig:0} shows the $\gamma$-ray and optical light curves. The \emph{Fermi}-LAT light 
curve is computed with an 8-weeks binning, integrating energies above 300~MeV. The time bin 
edges are chosen to conveniently overlap with the neutrino flare, highlighted by the green 
shaded area in \ref{fig:0}. Notably, the detection of the high-energy neutrino IC170922A 
(red line) coincides with the major $\gamma$-ray outburst experienced by TXS~0506+056 in 
the ten years of $\gamma$-ray monitoring, reaching a peak flux of 
$(1.20 \pm 0.07)\cdot 10^{-7}$cm$^{-2}$s$^{-1}$. The more recent 2018 data highlight 
the fading trend of the flare, while the flux remains still at higher-than-average values. 
The  behaviour in the optical band overall matches quite well the major $\gamma$-ray variations, 
displaying a lower-than average optical flux during times coincident with the neutrino flare. 
The brightest prolonged variations are coincident with the major $\gamma$-ray flare and IC170922A.

\section{Theoretical Setup}
\label{sec:Method}

\begin{figure}
\centering
  \includegraphics[width=0.48\textwidth]{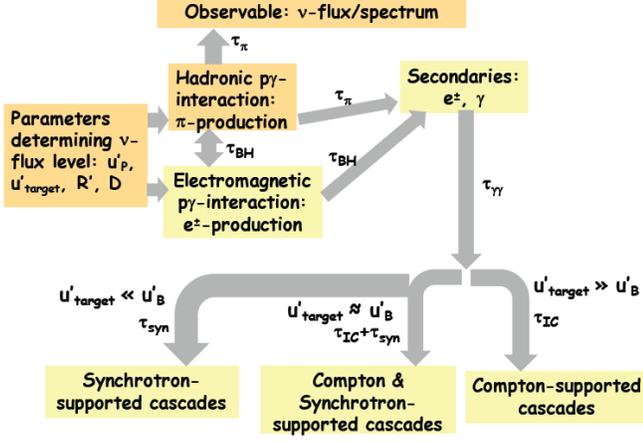}
\caption{Flow chart of our theoretical scheme. A fit to the IceCube neutrino spectrum during the
2014 -- 2015 neutrino flare determines a combination of parameters pertaining to the relativistic 
proton spectrum and the target photon field, leaving their relative normalization as a free parameter. 
A specific choice of proton power (and, thus, target photon energy density) determines the $\gamma\gamma$
opacity of the region. Depending on the relative energy densities of the magnetic field and the target 
photon fields, electromagnetic cascades develop in the synchrotron-, synchrotron-Compton, or Compton-dominated
regimes. }
\label{fig:scheme}   
\end{figure}

The following section is devoted to deriving the minimal requirements on the target photon field and
the relativistic proton population to explain the observed neutrino (flare) flux and spectrum 
in a way that is as independent 
of model assumptions as is possible. We will then discuss implications for 
the environment where the neutrino production occurs. The focus lies on the region where the neutrino 
production takes place. Our proposed procedure as illustrated by the flow chart in 
Fig.~\ref{fig:scheme} does not require a full multi-messenger modelling, 
but nevertheless will be able to constrain the family of emission models by comparing with MWL data.

In our minimal setup we consider all neutrinos to be produced photo-hadronically in the jet of 
TXS~0506+056. The calculations of all particle-photon and photon-photon interactions are carried 
out in the jet-frame, where we assume, for simplicity, the target radiation field to be isotropic.
For external target radiation fields the angular distribution of the target photons in the co-moving jet frame may become anisotropic and will lead to an increase or lowering of the particle-photon and photon-photon collision rate by an amount that depends essentially on the relative location of the emission region with respect to the source of the target photons \citep[e.g.,][]{ProtheroeDermer92,Dermer93,Roustazadeh10,Sitarek12,Dermer12}. We expect only minor changes of the corresponding photon emission for this case as discussed below (Sect.~\ref{sec:Cascading}).
A quantitative comprehensive 
study of general anisotropy effects for the present setup is in progress and will be presented in a forthcoming work (Reimer et al., in prep.).
The resulting 
yields are then transformed into the observer frame. The secondary particles from particle-photon 
interactions initiate pair cascades in the emission region which we follow to calculate the 
emerging photon SED. The pair cascading is considered to be linear i.e., 
the pair compactness is modest or low: $L'_i \sigma_T / 
(R' m_e c^3) < 10$, where $L'_i$ is the injected particle power, $\sigma_T$ the Thomson cross 
section, and $R'$ the size of the emission region \citep{Svensson87}, and the calculations 
consider the steady-state case.

\subsection{Proton-photon interactions and neutrino production}
\label{sec:NeutrinoProduction}

Neutrinos are produced through hadronic interactions of relativistic protons of (jet-frame) 
energy $E'_p = \gamma'_p \, m_p c^2$ with target photons of (jet-frame) energy $\epsilon'$ 
when the center-of-momentum frame energy $\sqrt{s}$ of the interaction is above threshold, 
i.e., $\sqrt{s} > \sqrt{s_{\rm thr}} = (m_p + m_{\pi})c^2 \simeq 1.08$~GeV. The  
nucleon energy required to produce neutrinos at hundreds of TeV energies, $E_{\nu} \equiv 100 \, E_{14}$~TeV 
is $E'_p \simeq 200 E_{14}/(D_{1} \xi_{0.05})$~TeV 
(i.e., $\gamma'_p = E'_p/m_p c^2 \simeq 2 \times 10^5 E_{14}/(D_{1} \xi_{0.05})$) 
where $\xi \equiv 0.05 \, \xi_{0.05}$ is the 
average neutrino energy per injected nucleon energy in photo-hadronic interactions \citep{Muecke99}. 
Protons 
of this energy are very inefficiently emitting synchrotron radiation in the magnetic 
fields typically assumed in hadronic blazar models ($B \sim 10$ -- $100$~G) with observed 
synchrotron cooling time scales of $t^{\rm obs}_{\rm psy} \sim 25 \, D_1^{-1} \, 
B_1^{-2} \, (\gamma / [6 \times 10^6])^{-1} \, {\rm yr}$, where $B_1 \equiv B / (10 \, {\rm G})$. 
The Larmor radius of protons with such energy is $r_L \sim 2 \times 10^{12} \, (\gamma / [6 \times 10^6]) 
\, B_1^{-1}$~cm, indicating that they are expected to be well confined within the emission 
region and can plausibly be accelerated by standard mechanisms, such as Diffusive Shock 
Acceleration. 

With the threshold condition, $\epsilon' E_p' \geq 0.14$~GeV$^2$, one finds the minimum energy of 
the target photons required to produce neutrinos at tens of TeV energies, $\epsilon'\geq 0.7 
D_{1} \xi_{0.05}/E_{14}$~keV. For pion (and neutrino) production at the $\Delta^+$ resonance, 
where $s = D_{\Delta^+}^2 = (1232 \, {\rm MeV})^2$,
target photons of $\epsilon' \geq 1.6 D_{1} \xi_{0.05}/E_{14}$~keV are required. 
In order to produce neutrinos in a broad spectral range we consider the 
injection of relativistic protons with a power-law distribution $dN'_p/dE'_p 
\propto E_p^{'-{\alpha_p}}$ in the range $E'_p=[m_p c^2, E'_{p,\rm max}]$ in a target 
photon field of photon energy density $u'_{\rm t}$, and use the SOPHIA Monte 
Carlo code \citep{SOPHIA} to calculate all resulting yields. We parametrize the 
target field using a simple power law for its differential photon spectrum 
$n'_{\rm t} \propto \epsilon'^{-\alpha_t}$ in the energy range $\epsilon'=[\epsilon'_{\rm min}, 
\epsilon'_{\rm max}]$ and re-construct (using the SOPHIA code) the {\it minimal target 
photon field}, irrespective of its origin, that is required to explain the observed 
neutrino spectrum upon injection of a power-law proton distribution. 
\footnote{The {\it minimal target photon field} is the narrowest possible soft photon spectrum 
with a shape and normalization that yields the observed 
neutrino spectrum upon injection of a power-law distribution of protons interacting photohadronically 
with these soft photons. The required normalization depends strongly on the proton distribution, 
unlike the photon spectral shape for sufficiently narrow target field spectra.}
 
\begin{figure}
\centering
  \includegraphics[height=6.5cm]{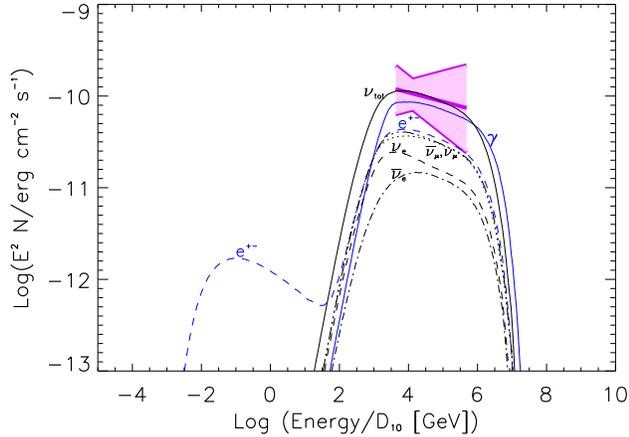}
\caption{Secondary particles (muon neutrinos: dotted black lines, anti-muon neutrinos: 
dashed-triple dotted black lines, electron neutrinos: dashed black lines, anti-electron 
neutrinos: dashed-dotted black lines, total neutrino yield: solid black line, 
electron-positrons: dashed blue line, photons: solid blue line) at production 
due to proton-photon interactions (photo-meson and Bethe-Heitler pair production) 
on the same target photon field, in comparison to the total flux as measured by 
IceCube during the 2014 -- 15 neutrino flare (violet bow tie), corrected for 
neutrino oscillations. Energies and fluxes are given in the observer frame.}
\label{fig:1}   
\end{figure}

\begin{figure}
\centering
  \includegraphics[height=6.5cm]{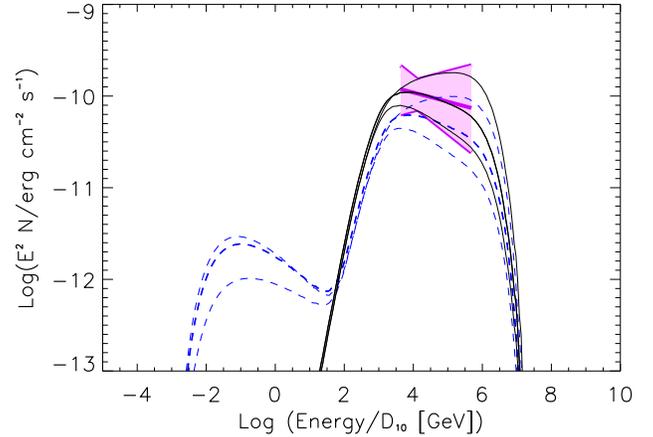}
\caption{Effect of varying the proton distribution: Total neutrino (solid black line) 
and electron-positron yields (dashed blue line) 
at production due to proton-photon interactions (photo-meson and Bethe-Heitler pair production) 
on the same target photon field as for Fig.~\ref{fig:1} for injection proton spectra with indices 
$\alpha_p =$1.8, 2.0, and 2.1, with a normalization that is higher (lower) by $\sim 10\%$ for the 
harder (softer) spectrum 
and $E'_{\rm p,max} = 30$~PeV to reflect the spread of the 
total neutrino flux as measured by IceCube during the 2014 -- 15 neutrino flare 
(violet bow tie), corrected for neutrino oscillations. Energies and fluxes are given in the observer frame.}
\label{fig:2}
\end{figure}

Fig.~\ref{fig:1} shows the resulting muon neutrino (dotted and dashed-triple-dotted black lines) and 
electron neutrino (dashed and dashed-dotted black lines) spectra at production for a 'blob' 
that is moving with $D=10$. The spectra are produced assuming that protons with a power-law 
spectrum with index $\alpha_p=2$ and energies up to $E'_{p,\rm max}=30$~PeV interact with a 
distribution of target photons with spectral index $\alpha_t=1$, between energies 
$\epsilon'_{\rm min}=10$~keV and $\epsilon'_{\rm max}=60$~keV. Neutron-decay neutrinos 
would appear at $\sim 10^{-3}$ smaller energies, and are not considered here. 
We adopted somewhat higher target photon energies than 
naively estimated so that the simulated neutrino spectrum matches the best-fit  spectrum 
for the 2014/15 neutrino flare reported by the IceCube collaboration. Looking at the 
SED of TXS~0506+056, the required energy of the target photon field coincides with the 
rising part of the second, high-energy hump, where the electromagnetic spectrum follows 
roughly $dN/dE\propto E^{-1}$. This motivates the 
choice of $\alpha_t=1$. The observational uncertainties of the neutrino spectrum are well 
captured by varying the proton spectral index, in the range $\alpha_p=2.0\pm^{0.1}_{0.2}$ 
 along with a corresponding proton spectrum normalization variation by $\pm 10\%$ 
(see Fig.~\ref{fig:2}). The calculation of the total neutrino spectrum (black solid lines in 
Fig.~\ref{fig:1} \& \ref{fig:2}) takes into account neutrino oscillations, and is compared to 
the IceCube observations (violet bow tie) of the neutrino flare. 
Once the target photon spectral shape and Doppler factor are fixed, 
the neutrino flux normalization depends further
on the injected proton energy density $u'_p$ (mildly coupled to $\alpha_p$), and, through the 
photomeson production 
opacity $\tau_{p\gamma}$, on the target photon energy density $u'_{t}$ and 
the size $R'$ of the emission region (see Fig.~\ref{fig:scheme}). The combination of these parameters is chosen 
to fit the observed neutrino flux (and hence is determined by observables), while 
each parameter individually is not required to be fixed. 
In case of target photon fields external to the jet particle-photon collisions may become anisotropic in the co-moving jet frame, which in turn affects the collision rates. The observed neutrino flux level would then be matched by adjusting the injected proton energy density correspondingly.

Secondary pairs are produced along with the neutrinos, and are depicted as the blue dashed-line bump
at high-energies ($\sim 0.1$~TeV-10~PeV) in Fig.~\ref{fig:1}, while the produced gamma-ray 
spectrum from meson decay is shown by the solid blue line. 
The ratio of the secondary pair and photon number density to that of the neutrinos remains constant, even in the case of scatterings involving target photon populations that are anisotropic in the jet frame.
Electromagnetic proton-photon interactions (Bethe-Heitler pair production) are 
guaranteed at sites where photo-hadronic interactions occur, and is therefore taken into 
account here (see Fig.~\ref{fig:scheme}). 

\begin{figure}
\centering
  \includegraphics[height=6.5cm]{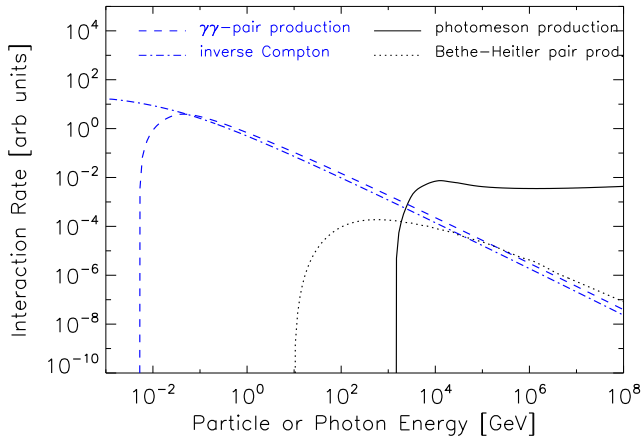}
\caption{Mean particle-photon interaction rates in the same target photon field as used in
Figs.~\ref{fig:1} and \ref{fig:2}. Photo-meson production: solid black line, Bethe-Heitler 
pair production: dashed black line, inverse Compton scattering: dashed blue line, photon-photon 
pair production: dotted blue line. All energies are given in the jet frame.}
\label{fig:3}
\end{figure}

We simulate Bethe-Heitler 
pair production in the same target photon field using the Monte Carlo code of 
\citet{Protheroe96}, and using the same proton injection distribution and 
flux normalization. The resulting pair yields upon production are shown as 
the blue dashed line lower-energy (10 MeV - 30 GeV) hump in Fig.~\ref{fig:1} for a Doppler factor $D=10$ 
as an example, and its spread due to the observational uncertainties of the neutrino 
spectrum is depicted in Fig.~\ref{fig:2}. Note that the ratio of the pair-to-pion production 
rates in the same target photon field (see Fig.~\ref{fig:3}) is fixed by the nature of the 
interaction.

Fig.~\ref{fig:3} depicts interaction rates for various
processes. It illustrates the dominance of losses due to Bethe-Heitler pair production for low 
proton energies while above the threshold for meson production photo-hadronic interactions 
dominate. While the minimal energy range of the target photon field is fixed to some extent by the 
energy range of the observed neutrino spectrum (modulo the Doppler factor in jetted neutrino 
sources) its shape is less constrained. Indeed, by choosing softer (e.g. $\alpha_t=2-3$) or 
harder (e.g. $\alpha_t=0-0.5$) power-law target photon fields we were able to find equally 
good representations of the observed neutrino spectrum, with only very small adjustments of 
the remaining parameters (e.g., $\alpha_p$, etc). We attribute this behaviour to the narrowness 
of the observed neutrino spectrum. Changing the target photon spectral index in the considered 
narrow energy range is expected to cause only minor changes in the associated pair cascade 
spectrum (see Sect.~\ref{sec:Cascading}), well within the uncertainties implied by the observed 
neutrino spectrum. The situation changes somewhat when extending sufficiently peaked target 
photon spectra to a broader energy range. For example, by extending the $\alpha_t=1$ target 
photon spectrum down to $\epsilon'_{\rm min}=1$~eV, we were also able to find a representation 
of the observed neutrino spectrum for $\alpha_p=2.2$ and $E'_{p,\rm max}=20$~PeV. While the 
corresponding impact on the synchrotron-supported cascades turns out minor (except for a 
broadening of the absorption troughs), the Compton-supported 
cascades will become broader (see Sect.~\ref{sec:ICcascade}).

Finally, varying the Doppler factor shifts the energy range of the minimal target photon field 
and injected proton spectrum correspondingly, while the required flux normalization to reach 
the observed neutrino flux level can be adjusted by changing the injected proton energy density 
accordingly. We scanned from $D=1$ (suitable for misaligned jetted AGN) up to $D=50$ (suggested 
by minute-time scale gamma-ray flux variations found in some blazars \citep[e.g.,][]{Begelman08} 
to find satisfactory representations of the observed neutrino spectrum by using 
$\epsilon'_{\rm min} = 1 \ldots 170$~keV,  $\epsilon'_{\rm max} = 4\ldots 1000$~keV 
and $E'_{\rm p,max} = 3\ldots 300$~PeV.

To summarize, Fig.~\ref{fig:1} shows the distribution of all secondary pairs and $\gamma$-rays 
that are inevitably produced along with the (observed) neutrinos. By varying the proton spectral 
index $\alpha_p=2.0\pm^{0.1}_{0.2}$ we propagate the observed uncertainties of the neutrino 
spectrum to the corresponding pair and $\gamma$-ray distributions.
These pairs and $\gamma$-rays typically initiate electromagnetic cascades 
(see Fig.~\ref{fig:scheme}) which re-distribute 
their power to energies where the produced photons can eventually escape the source. 
The $\gamma\gamma$-pair production optical depth and its energy-dependence determines 
the spectral escape probability of the photons, and is therefore central for calculating 
the emerging photon spectrum associated with the neutrino spectrum.

\subsection{Pair cascading}
\label{sec:Cascading}

The high-energy pairs and $\gamma$-rays that were produced along with the neutrino flux in the 
minimal target radiation field are injected into a pair cascading code.
In a radiative and magnetized environment the electromagnetic cascade develops rapidly in 
the target photon field through photon-photon pair production, and is either driven by mainly 
inverse Compton scattering in the same radiation field ('Compton-supported cascades') if 
$u'_{\rm t}\gg u'_B$ (with $u'_B$ the magnetic field energy density), by synchrotron 
radiation ('synchrotron-supported cascades') if $u'_{\rm t}\ll u'_B$, or by both 
('synchrotron-Compton cascades') in cases where the magnetic field energy density turns 
out to be comparable to the target photon energy density (see Fig.~\ref{fig:scheme}). 
Once the cascade photons reach 
down to energies too low for pair production, the cascade development ceases, and the 
remaining pairs lose their energy via inverse Compton and/or synchrotron emission. The 
emerging photon spectrum is governed by both, the energy-dependent photon-photon pair 
production optical depth, and the dominating radiative dissipation process.

The pair cascading code employs the matrix multiplication method described by \citet{Protheroe93}
and \citet{Protheroe96}. Monte Carlo programs for photon-photon pair production and inverse 
Compton scattering are used to calculate the mean interaction rates (see Fig.~\ref{fig:3}) and the 
secondary particle and photon yields due to the interaction with the target radiation field. 
Note that inverse Compton scattering during the cascading proceeds in the Klein-Nishina 
regime, and transitions to the Thomson regime only below the threshold for pair production. 
The calculation of the synchrotron yields follows \citet{Protheroe90} \citep[see also][]{Pacholczyk70}. 
The yields are then used to build-up transfer matrices which describe the change in the electron 
and photon spectra after propagating a given time step $\delta t$, which we chose
as the dynamical time scale of the problem. 
The escape probability for photons is calculated using the formula given in 
\citet{osterbrock} for a spherical region, while charged particles are assumed not to escape. 
For steady-state spectra we continue the transfer process until convergence is reached.
In each time step energy conservation is verified.

In the following, all the photon models are generated with a normalization that yields the 
observed neutrino flux.

\subsubsection{Compton-supported Cascades}
\label{sec:ICcascade} 

\begin{figure}
\centering
  \includegraphics[height=6.5cm]{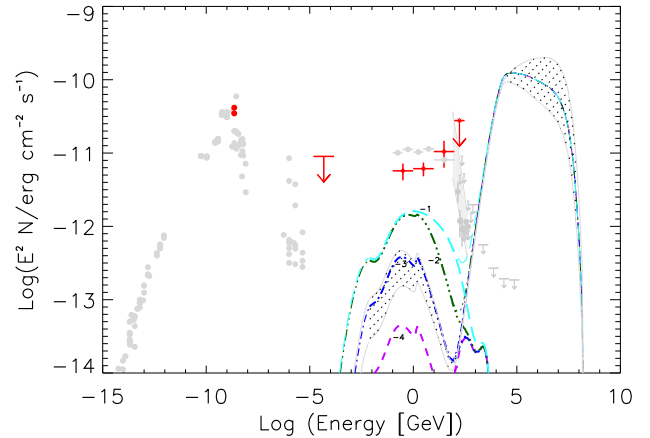}
\caption{Compton-supported cascade spectra arriving at Earth without (thin lines) and including 
absorption in the EBL (thick lines) using the model of \citet{Franceschini17} for 
$\log(\tau_{\gamma\gamma,\rm max})= -4$ (short dashed line),  $-3$ 
(dashed-dotted line),  $-2$ (dashed-triple-dotted line),  $-1$ (long dashed line)
as indicated, and $D=10$. The shaded areas represent the spread of the cascade spectra, for the 
case $\tau_{\gamma\gamma,\rm max}=10^{-3}$ as an example, propagated from the uncertainties of 
the observed neutrino spectrum.
The red data points (ASAS-SN, {\it SWIFT-BAT}, {\it Fermi}-LAT) 
depict the quasi-simultaneous observations while the grey data points represent archival data 
\citep{IceCube1}. VHE data (all MAGIC data from \cite{MAGIC}; all VERITAS data from \cite{VERITAS}, 
HAWC archival data from \cite{IceCube1} are not simultaneous to the 2014/15 neutrino flare and 
are included in grey.}
\label{fig:4}
\end{figure}

\begin{figure}
\centering
  \includegraphics[height=6.5cm]{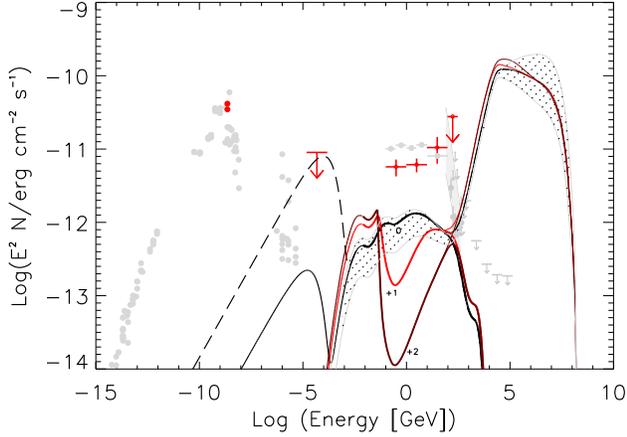}
\caption{Compton-supported cascade spectra arriving at Earth without (thin lines) and including 
absorption in the EBL (thick lines) for $\log(\tau_{\gamma\gamma,\rm max})=0, 1, 2$ as indicated,
and $D=10$. The shaded areas represent the spread of the cascade spectra, for the 
case $\tau_{\gamma\gamma,\rm max}=1$ as an example, propagated from the uncertainties of 
the observed neutrino spectrum. For the $\tau_{\gamma\gamma,\rm max}=1$-case we have also 
added the corresponding proton synchrotron radiation component for two 
sub-equipartion ($u'_B\ll u'_t$) field strengths: 0.5~G (solid line), 3~G (dashed line). 
The data points are the same as shown in Fig.~\ref{fig:4}.}
\label{fig:5}
\end{figure}

\begin{figure}
\centering
  \includegraphics[height=6.5cm]{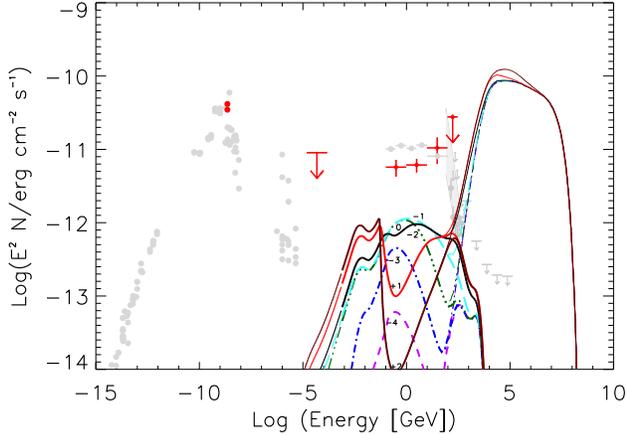}
\caption{Compton-supported cascade spectra arriving at Earth without (thin lines) and including 
absorption in the EBL (thick lines) for 
$\tau_{\gamma\gamma,\rm max}=10^{-4}$ (short dashed line),  $10^{-3}$ 
(dashed-dotted line),  $10^{-2}$ (dashed-triple-dotted line),  $10^{-1}$ (long dashed line), 
1, 10, 100 (solid lines) as indicated, and $D=1$. 
The data points are the 
same as shown in Fig.~\ref{fig:4}.}
\label{fig:6} 
\end{figure}

\begin{figure}
\centering
  \includegraphics[height=6.5cm]{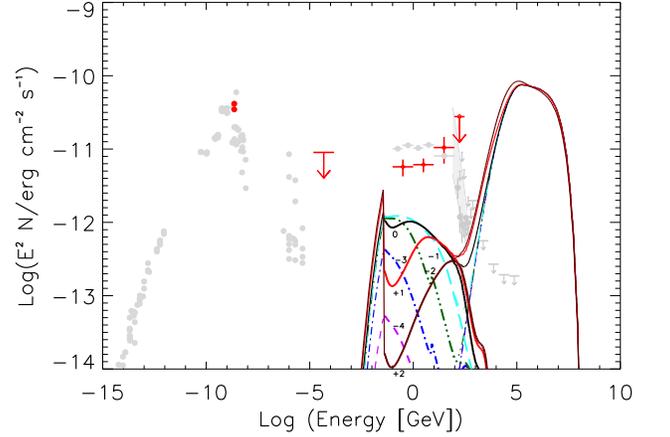}
\caption{Compton-supported cascade spectra arriving at Earth without (thin lines) and 
including absorption in the EBL (thick lines) for 
$\tau_{\gamma\gamma,\rm max}=-4$ (short dashed line),  
$-3$ (dashed-dotted line),  $-2$ (dashed-triple-dotted line),  $-1$ 
(long dashed line), 0, 1, 2 (solid lines) as indicated, and $D=50$. 
The data points are the same as shown in Fig.~\ref{fig:4}.}
\label{fig:7}       
\end{figure}

If inverse Compton scattering supports the cascade ($u'_{t}\gg u'_B$), the emerging 
photon spectra are fully determined once the opacity due to photon-photon pair production is fixed,
because $\tau_{\gamma\gamma}\propto \tau_{\rm IC}$ (with $\tau_{\rm IC}$ the inverse Compton optical depth). This energy-dependent opacity in turn is related 
to the optical depth of photon-proton interactions, which itself is linked to the neutrino spectral 
flux (see Sect.~\ref{sec:NeutrinoProduction}). Compton-supported cascading on the minimal target photon 
field (as determined in Sect. 3.1)
therefore provides the corresponding minimal cascade 
flux for $u'_{t}\gg u'_B$. In 
the following we choose to vary the maximum of the energy-dependent photon-photon opacity, 
$\tau_{\gamma\gamma,\rm max}$ (which here serves as a proxy for ($R'\cdot u'_{t}$)) 
to calculate the corresponding cascade spectra. Figs.~\ref{fig:4} -- \ref{fig:7} show our 
results for Doppler factors $D=1, 10$ and $50$ and $\log(\tau_{\gamma\gamma,\rm max})= -6, -5, 
\ldots, 0, 1, 2$, covering the optically thin and thick cases, and compares them to the MWL 
SED of TXS~0506+056 during the period of the neutrino flare. The shaded areas represent the 
spread of the cascade spectra (shown for selected $\tau_{\gamma\gamma,\rm max}=10^{-5}, 1, 100$ 
as examples) due to the uncertainties in the observed neutrino spectrum.

The cascade spectra at source (thin lines) are then corrected for absorption in the extragalactic 
background light (EBL) using the model of \citet{Franceschini17}. EBL-corrected cascade spectra 
are shown as thick lines. For $\tau_{\gamma\gamma,\rm max}\ll 1$ (e.g., Fig.~\ref{fig:4}), 
internal absorption can be neglected, and the emerging inverse Compton flux increases with 
increasing opacity since $\tau_{\gamma\gamma}\propto \tau_{\rm IC}$. None of these cascade 
spectra reach the flux level of the MWL (in particular, {\it Fermi}-LAT) data. Therefore,
in these cases, most of the photon flux has to be produced by emission processes occuring 
in the source that were not initiated by proton-photon interactions. When increasing the 
opacity further, $\tau_{\gamma\gamma}\gg 1$ (e.g., Fig.~\ref{fig:5}), deep absorption 
troughs in the GeV-range appear. Any photons in the GeV range produced co-spatially to the 
neutrino flux, will inevitably suffer from internal absorption and will subsequently be
cascaded to lower energies, independent of their production process. In this case of an 
efficient neutrino producer, the observed GeV-flux has to be produced either in a region 
within the jet of TXS~0506+056 that has no dense radiation fields at keV energies, or in 
a different source altogether. As a consequence, there is no causal connection between the 
observed neutrino flux and observed GeV-flux. These findings are similar for all values 
of the Doppler factor inspected (see Figs.~8-9).

\begin{figure}
\centering
  \includegraphics[height=6.5cm]{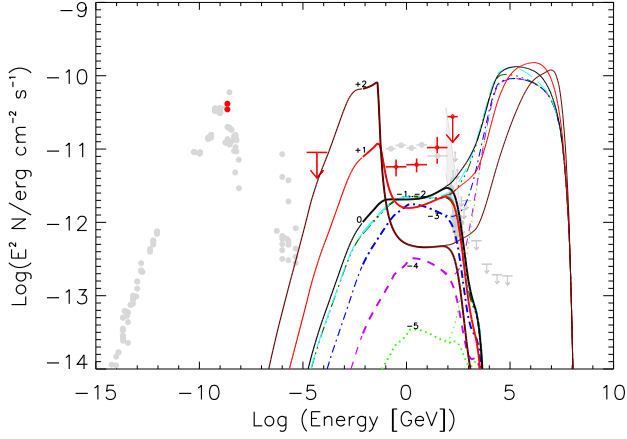}
\caption{Spectra from Compton-supported cascading in a broader target photon field 
($\epsilon'_{\rm min}=1$eV, $\epsilon'_{\rm max}=60$keV, $\alpha_t=1$) arriving at 
Earth without (thin lines) and including absorption in the EBL (thick lines) for 
$\log(\tau_{\gamma\gamma,\rm max})=-5$ (dotted line),  $-4$ 
(short dashed line),  $-3$ (dashed-dotted line), 
$-2$ (dashed-triple-dotted line),  $-1$ (long dashed line), 0, 1, 2 (solid lines) 
as indicated, and $D=10$. The data points are the same as shown in Fig.~\ref{fig:4}.}
\label{fig:8}       
\end{figure}

As pointed out in Sect.~\ref{sec:NeutrinoProduction} broader target radiation fields may also 
lead to photo-hadronically produced neutrino spectra that fit the observations (e.g., extending 
the target field $n'_t\propto \epsilon^{'-1}$ to lower energies, e.g., $\epsilon'_{\rm min}=1$eV). 
Inspecting the associated Compton-supported cascade spectra shows broader SEDs than for the case of 
narrow target photon fields (see Fig.~\ref{fig:8}). This is because the photon energy associated with 
the lowest-energy radiating electron is lower for $\epsilon'_{\rm min}=1$~eV (broad 
target field) than for $\epsilon'_{\rm min}=10$~keV (narrow target field). In addition, 
the overall cascade flux increases notably below the GeV regime when increasing the energy 
range of the target photon 
field. We note that in such cases X-ray data may be able to place stringent constraints on 
models.

The Compton optical depth $\tau_{\rm{IC}}$ may be altered when scatterings sample photon fields that are anisotropic in the co-moving frame. Since  $\tau_{\rm{IC}}$ is directly linked to $\tau_{\gamma\gamma}$ (see Fig.~\ref{fig:3}) and the emerging cascade spectra presented here are proxied by $\tau_{\gamma\gamma}$, no significant changes of the Compton cascade spectra for a given $\tau_{\gamma\gamma}$ are expected in such cases.

\subsubsection{Synchrotron-supported Cascades}
\label{sec:SynCascade}

If the magnetic energy density $u'_B$ is much larger than the target photon field 
energy density $u'_{\rm t}$ the electromagnetic cascades are supported by 
synchrotron radiation. In blazar jet environments the magnetic field strength 
$B$ (assumed constant here) is typically $\ll B_{\rm cr}=2\pi m_e^2 c^3/(eh)
= 4.4 \times 10^{13}$~G, the critical magnetic field strength for radiating 
pairs, which is therefore also our assumption here. 
\begin{figure}
\centering
  \includegraphics[height=6.5cm]{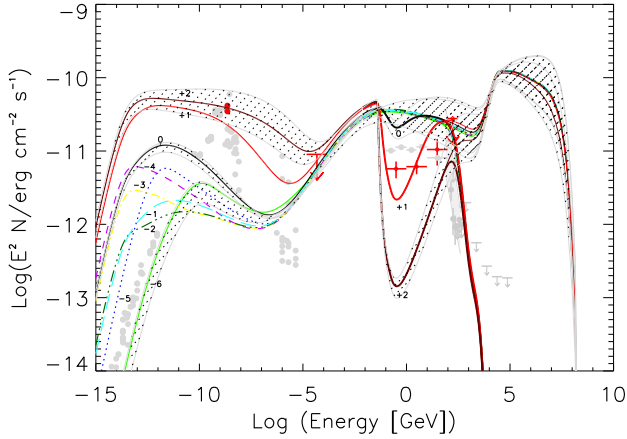}
\caption{Synchrotron-supported cascade spectra arriving at Earth without 
(thin lines) and including absorption in the EBL (thick lines) for 
$u_B=10u'_{\rm target}$ and $\log(\tau_{\gamma\gamma,\rm max})=-6$ 
(solid line), $-5$ (dotted line),  
$-4$ (short dashed line),  $-3$ (dashed-dotted line),  
$-2$ (dashed-triple-dotted line),  $-1$ (long dashed line), 
0, 1, 2 (solid lines) as indicated, and $D=10$. The shaded areas 
represent the spread of the cascade spectra, for the cases 
$\tau_{\gamma\gamma,\rm max}=10^{-6}, 1, 100$ as examples, propagated 
from the uncertainties of the observed neutrino spectrum. The data 
points are the same as shown in Fig.~\ref{fig:4}.}
\label{fig:9}  
\end{figure}
\begin{figure}
\centering
  \includegraphics[height=6.5cm]{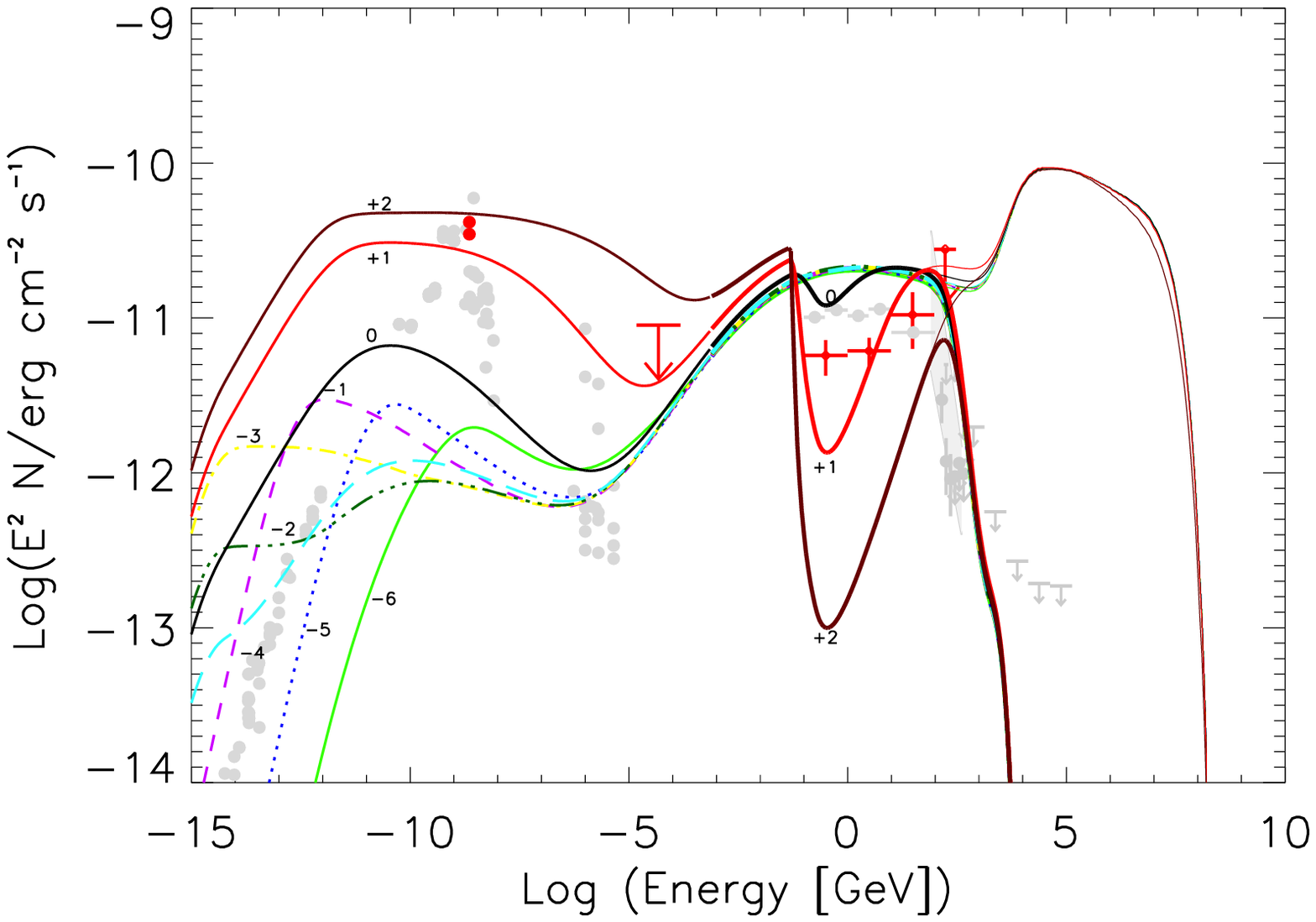}
\caption{Same as Fig.~\ref{fig:9} except $D=1$.}
\label{fig:10} 
\end{figure}
\begin{figure}
\centering
  \includegraphics[height=6.5cm]{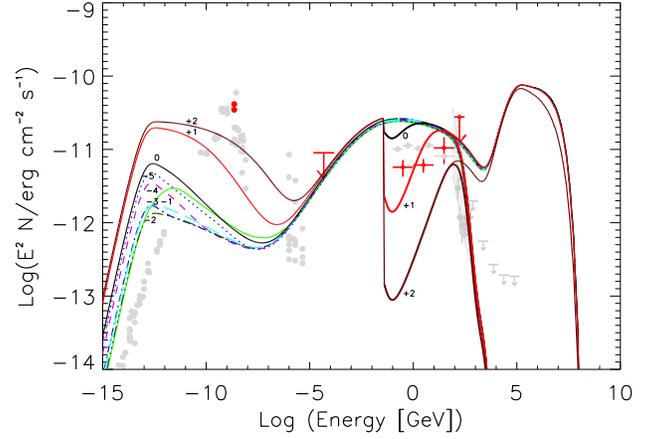}
\caption{Same as Fig.~\ref{fig:9} except $D=50$.}
\label{fig:11}
\end{figure}
As an example we show in Figs.~\ref{fig:9} -- \ref{fig:11} the case where 
$u'_B = 10 \, u'_{\rm t}$ with the above specified target photon spectrum, 
again corrected for absorption in the EBL. Synchrotron-self absorption is not 
taken into account. The uncertainties in the IceCube-flux are propagated again 
to the corresponding cascade spectra, and depicted in Figs.~\ref{fig:9} -- \ref{fig:11} 
as shaded areas for the cases $\tau_{\gamma\gamma,\rm {max}}=10^{-6}, 1, 100$. 
In the optically thin case the standard synchrotron cooled spectra for a two-component 
(meson-decay and the lower energy Bethe-Heitler pairs) pair population is recovered,
with increasing cooling for increasing field strength (see Fig.~\ref{fig:12}).  Note 
that for $B \ll B_{\rm cr}$ the energy loss rate is $\propto \gamma_e^2$ for all energies 
whereas for Compton-loss dominated pair cascades the interaction probability 
changes from $\propto \gamma_e$ in the Klein-Nishina regime to $\propto \gamma_e^2$ 
in the Thomson regime. This is reflected in the resulting shape of the cascade 
spectra. 
Because the secondary pair distribution resulting from proton-photon interactions is fixed by the observed neutrino
spectrum the corresponding synchrotron-supported cascade emission in the optically thin case is expected to be unaffected by a possible anisotropy of the collisions.
In the optically thick case, $\tau_{\gamma\gamma}\gg 1$, the number of 
pairs, produced from synchrotron photons, increases dramatically in the cascade. 
The amount of reprocessing is given by $\tau_{\gamma\gamma}$, and the resulting cascade flux
for a given $\tau_{\gamma\gamma}$ is therefore independent of whether it is produced by isotropic or anisotropic scatterings,
as long as the contribution to the scattering rate per unit scattering angle is energy-independent.
Also, the produced pairs are more energetic in environments with high magnetic 
fields than in sites of low magnetization (see Fig.~\ref{fig:13}). Similarly as 
for the optically thick Compton-supported cascades we note deep absorption troughs 
at GeV energies leading to the same implications for the association between the 
neutrino and GeV-source, i.e., the lack of a causal connection between the 
observed neutrino and GeV-flux.
Note that the hump-like feature at high energies ($\sim 1$GeV-1TeV) in the optically 
thick cases disappears by
sufficiently broadening the target photon spectrum to lower energies.
\begin{figure}
\centering
  \includegraphics[height=6.5cm]{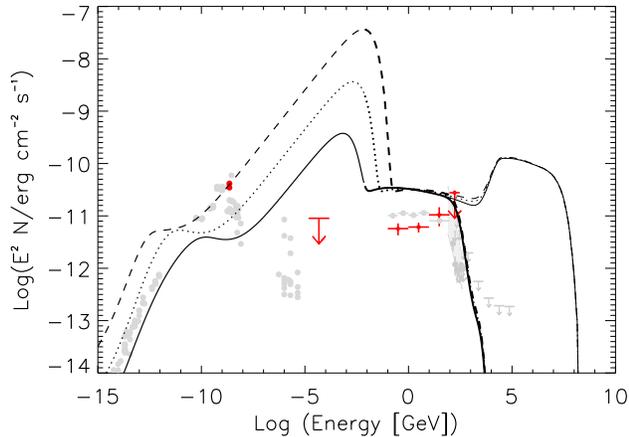}
\caption{Synchrotron-supported cascade spectra, including the proton synchrotron radiation components, 
arriving at Earth without (thin lines) and including absorption in the EBL (thick lines) for 
$u_B'=10u'_{\rm t}$ 
(solid line), $u_B'=10^2u'_{\rm t}$ (dotted line), $u_B'=10^3u'_{\rm t}$ 
(dashed line),  $\tau_{\gamma\gamma,\rm max}=10^{-6}$, and $D=10$. 
The data points are the same as shown in Fig.~\ref{fig:4}.}
\label{fig:12}
\end{figure}
\begin{figure}
\centering
  \includegraphics[height=6.5cm]{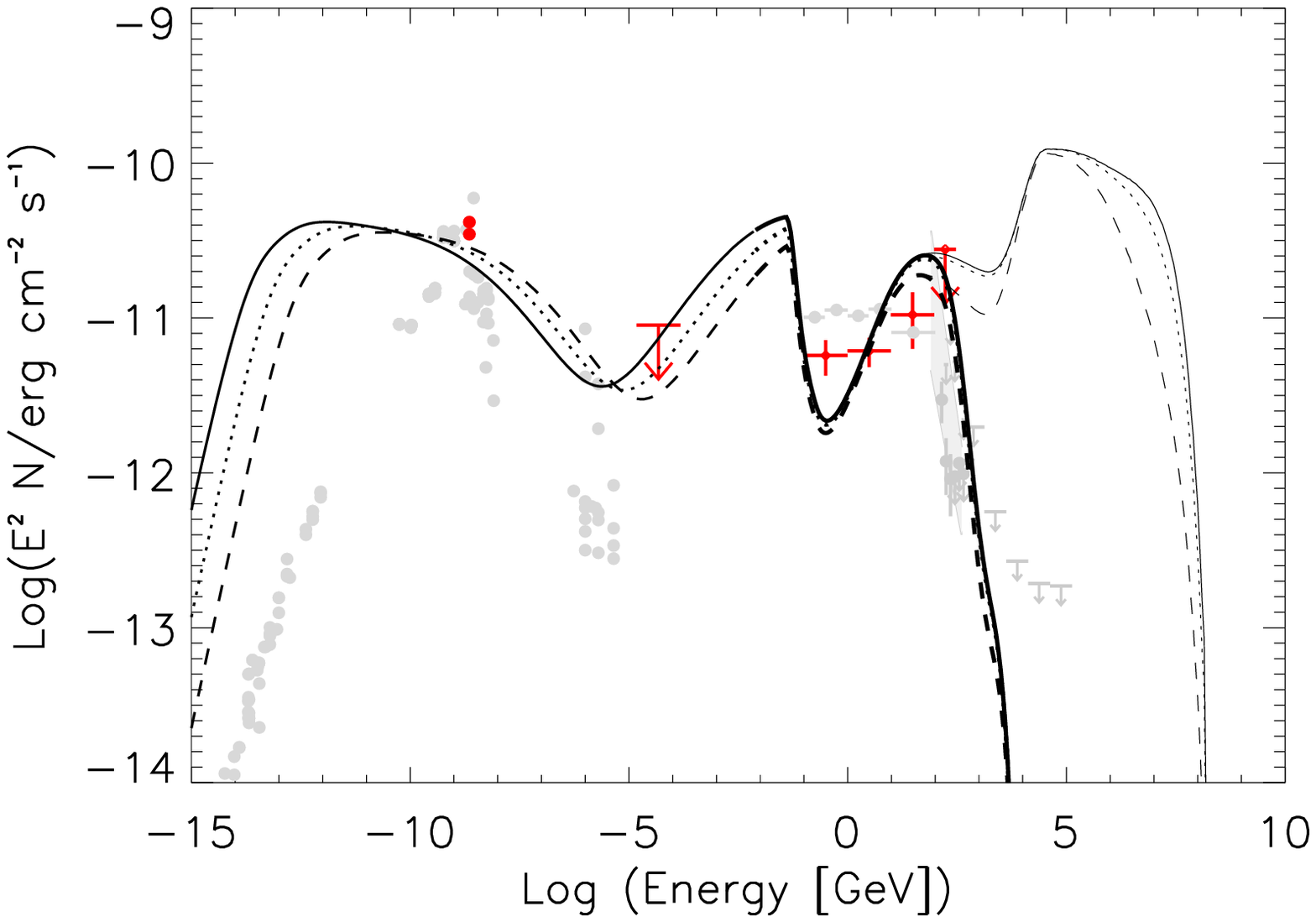}
\caption{Synchrotron-supported cascade spectra arriving at Earth without (thin lines) and 
including absorption in the EBL (thick lines) for $u_B'=10u'_{\rm t}$ (solid line), 
$u_B'=10^2u'_{\rm t}$ (dotted line), $u_B'=10^3u'_{\rm t}$ (dashed line), 
$\tau_{\gamma\gamma,\rm max}=10$, and $D=10$. The data points are the same as 
shown in Fig.~\ref{fig:4}.}
\label{fig:13} 
\end{figure}
As compared to the Compton-supported case, the synchrotron-supported cascades 
extend down to much lower energies: whereas the inverse Compton photon spectrum 
extends down to $\propto \gamma^2_{min} \epsilon'_{t,\rm min}$, for the synchrotron 
spectrum the lowest energy reached is $\propto \gamma^2_{min}(B/B_{\rm cr}) 
m_e c^2$ with typically $(B/B_{\rm cr}) m_e c^2 \ll \epsilon'_{t,\rm min}$ 
for an observed neutrino spectrum extending down to tens of TeV.

When compared to the quasi-simultaneous observed SED we note that deep 
optical-to-X-ray observations are able to provide sensitive constraints, 
in particular on the efficiency of neutrino production. For the present 
case we find that the {\it BAT} upper limit constrains the neutrino-producing 
environment to $\tau_{\gamma\gamma, \rm max}<100$ for $D\leq 10$, whereas 
the {\it Fermi}-LAT spectrum excludes $\tau_{\gamma\gamma, \rm max}<10$.

\subsubsection{Synchrotron-Compton Cascades}
\label{sec:ICsynCascade}

\begin{figure}
\centering
  \includegraphics[height=6.5cm]{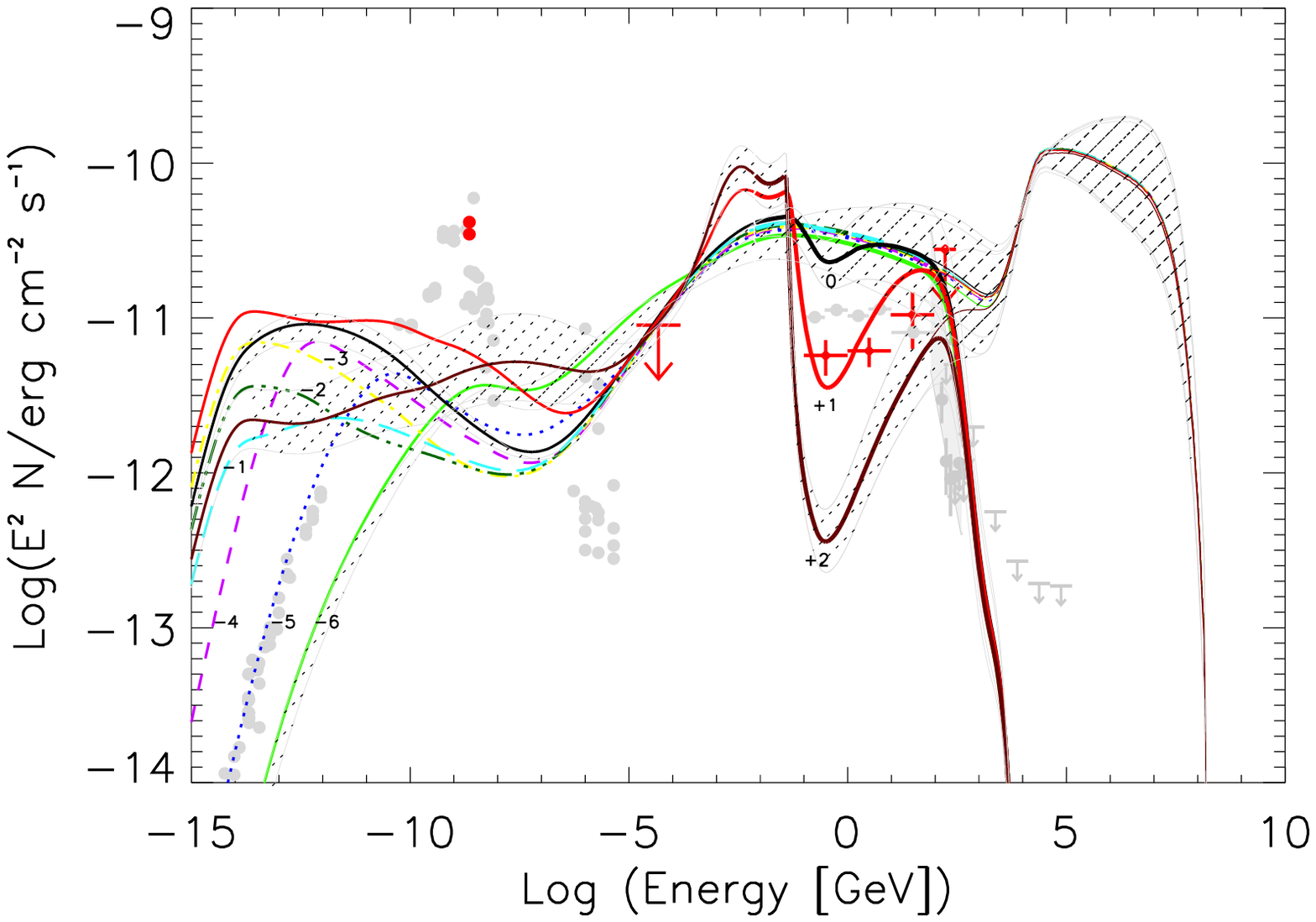}
\caption{Synchrotron-Compton cascade spectra arriving at Earth without (thin lines) and 
including absorption in the EBL (thick lines) for $u_B=u'_{\rm target}$ and 
$\tau_{\gamma\gamma,\rm max}=10^{-6}$ (solid line), $10^{-5}$ (dotted line),  
$10^{-4}$ (short dashed line),  $10^{-3}$ (dashed-dotted line),  
$10^{-2}$ (dashed-triple-dotted line),  $10^{-1}$ (long dashed line), 1, 10, 100 (solid lines) 
as indicated, and $D=10$. The shaded areas represent the spread of the cascade spectra, for the 
cases $\tau_{\gamma\gamma,\rm max}=10^{-6}, 1, 100$ as examples, propagated from the uncertainties 
of the observed neutrino spectrum. The data points are the same as shown in Fig.~\ref{fig:4}.}
\label{fig:14} 
\end{figure}

\begin{figure}
\centering
  \includegraphics[height=6.5cm]{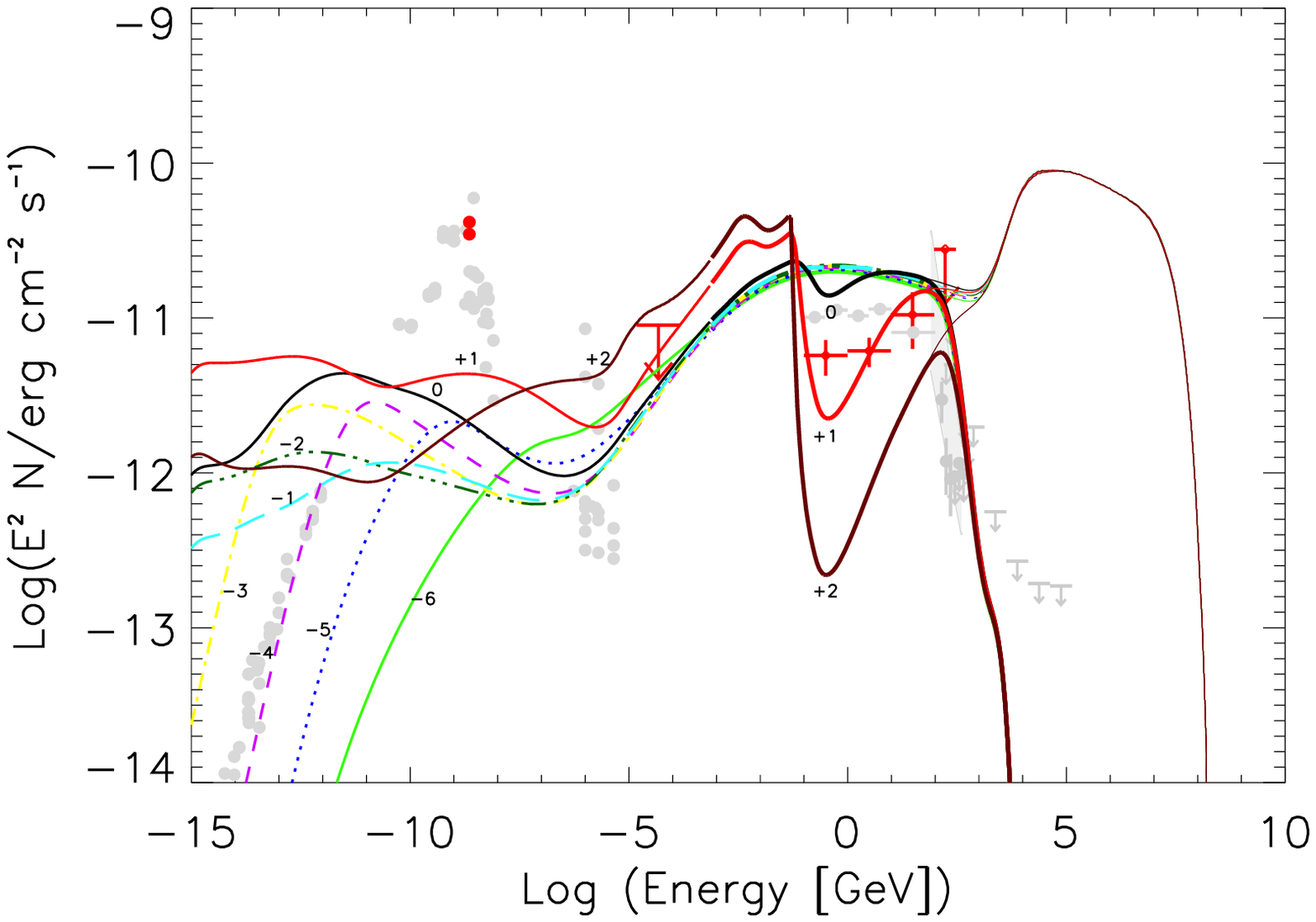}
\caption{Same as Fig.~\ref{fig:14} except $D=1$.}
\label{fig:15} 
\end{figure}

\begin{figure}
\centering
  \includegraphics[height=6.5cm]{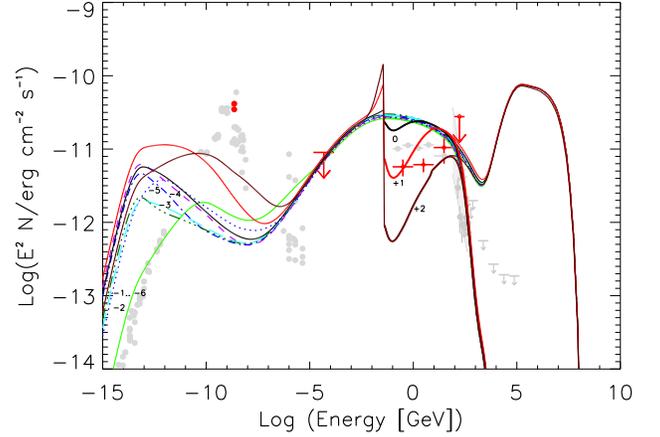}
\caption{Same as Fig.~\ref{fig:14} except $D=50$.}
\label{fig:16} 
\end{figure}

Generally in environments where $u'_B \sim u'_{\rm t}$ both, synchrotron as well as 
Compton losses have to be considered when following electromagnetic cascades. 
In the high-energy regime, down to the threshold for pair production, the 
synchrotron-Compton (see Figs.~\ref{fig:14}
-- \ref{fig:16}) and synchrotron-supported cascade SEDs are very similar. This is because 
Compton-scattering proceeds in the Klein-Nishina regime here at a rate much smaller than 
the synchrotron loss rate, and the resulting spectra are determined by synchrotron radiation 
and pair production.
At lower energies, photon production takes 
place through synchrotron and inverse Compton scattering on equal footing, which is reflected 
in the resulting cascade spectra.
With increasing opacity the number of (cascade) pairs rises. At MeV-energies the resulting SED is 
dominated by inverse Compton photons from the primary Bethe-Heitler pairs and cascade pairs, and 
synchrotron photons
from the $\pi^\pm$-decay pairs, while the SED at lower einergies is due to synchrotron radiation from the
Bethe-Heitler and cascade pairs.

By comparing also here with the quasi-simultaneous observed SED we find again the {\it BAT} upper 
limit to place constraints on the neutrino production efficiency $\tau_{p\gamma}$: Only
$\tau_{\gamma\gamma,\rm max} \sim$ a few tens for small Doppler factors $D<10$, 
$\tau_{\gamma\gamma,\rm max} \sim 10$ if $D=10$, and $\tau_{\gamma\gamma,\rm max} \sim$ 
a few hundreds for large $D\gg 10$ do not violate the minimal cascade fluxes computed, 
similar as for the synchrotron-supported cascades.

\subsection{\label{sec:absorption}On efficient neutrino production}

Since the cross section maximum for photon-photon pair production (near its threshold) 
compares to the maximum cross section value for photomeson production (at the $\Delta_{1232}$-resonance 
near threshold) as $\sigma_{\gamma\gamma,{\rm max}} \approx 300 \, \sigma_{p\gamma,{\rm max}}$, we 
expect in typical AGN environments the maximum optical depth of the respective interactions in the 
same target photon field to approximately scale correspondingly. In particular, for the case of an 
efficient neutrino producer, i.e., $\tau_{p\gamma} > 1$, one yields $\tau_{\gamma\gamma} \gg 1$, 
and deep absorption troughs appear in the 
SED at energies $E'_\gamma$ where $\gamma\gamma$-pair production sets in (see also discussions 
in, e.g., \cite{Begelman90,Dermer07,Murase16}):

\begin{equation}
E'_{\gamma, \rm thr} = \frac{s_{\gamma\gamma,\rm thr}}{2\epsilon' (1-\cos\theta)} = 
\frac{4m_e^2c^4}{2\epsilon' (1-\cos\theta)}
\label{ggthreshold}
\end{equation}
where $\sqrt{s_{\gamma\gamma,\rm thr}}$ is the center-of-momentum-frame threshold energy of the interaction, $\epsilon'$ 
the target photon energy and $\theta$ the interaction angle. Considering that neutrinos are produced in the same target 
photon field via photomeson production dominantly in the $\Delta$-resonance region, the energy of the target photons is 
related to the observed neutrino energy $E_{\nu,\rm obs}$ through
\begin{equation}
\epsilon'\gtrsim \frac{s_\Delta-m_p^2c^4}{2E'_p (1-\cos\theta)} = \frac{(s_\Delta-m_p^2c^4)\xi D}{2E_{\nu,\rm obs} (1-\cos\theta)}
\label{pgthreshold}
\end{equation}
with $s_\Delta\simeq$(1.232~GeV)$^2$ and assuming $\beta_p=1$. Note that the interaction angle for proton-photon and 
photon-photon interactions is in this case expected to be similar for highly relativistic protons.
By combining equations (\ref{ggthreshold}) and (\ref{pgthreshold}) one can then relate the (observer frame) 
photon energy $E_{\gamma,\rm obs}$ to the detected neutrino energy through
\begin{equation}
E_{\gamma,\rm obs} \lesssim \frac{4 m_e^2 c^4 E_{\nu,\rm obs}}{(s_\Delta-m_p^2c^4)\xi}\approx 3.3 \cdot 10^{-5} E_{\nu,\rm obs} \xi_{0.05}^{-1}\,.
\label{egobs}
\end{equation}
If neutrinos are very efficiently produced photohadronically, the associated cascade photons therefore can only escape the source
at energies $\lesssim E_{\gamma,\rm obs}$. For a neutrino flare spectrum in the 30~TeV -- 3~PeV range, the corresponding cascade 
spectrum is hence expected at $\ll 1$~GeV. This applies also to $>$~GeV photons from alternative radiation processes.
As a consequence, one does not expect a causal connection between the GeV-flux detected by the LAT from TXS~0506+056 and the 
observed 'flare' IceCube neutrinos for an efficient neutrino producer, irrespective of the dominant dissipation process.

\section{Implications on the jet power and target photon fields}
\label{sec:Implications}

We now derive constraints on the jet power using the observed neutrino flux in combination with 
limits for their production efficiency, derived via the minimal cascading flux (see Sect.~\ref{sec:Cascading}). 
This can then be used, in a subsequent step, to deduce limits on the origin of the target photon field.
Here, we will distinguish two possible scenarios, which can be thought of as extreme, limiting cases: 
(a) a target photon field that 
is co-moving with the emission region (such as the electron-synchrotron emission), or (b) a stationary 
target photon field in the AGN rest frame. In case (a), the target photon energy $\epsilon'_t$ corresponds 
to an observed photon energy of $\epsilon_{\rm obs} \geq 16 D_{1}^2 \xi_{0.05}/E_{14}$~keV (i.e., 
hard X-rays), while in case (b), the external (stationary) target photon field is Doppler boosted 
into the blob frame, so that $\epsilon_{\rm obs} \geq 0.16 \xi_{0.05}/E_{14}$~keV (i.e., UV-to-soft X-rays).

\subsection{Target Photon Energy Density and Proton Kinetic Power}
\label{sec:power}

As in Section \ref{sec:Method} we consider a proton spectrum of the form $N_p (\gamma_p) = N_0
\, \gamma_p^{-\alpha_p}$ with $\alpha_p = 2$. The co-moving neutrino luminosity (see 
Sect.~\ref{sec:NeutrinoData}) can be related to the proton energy loss rate due to photopion 
production, given by \citet{KA08}
\begin{equation}
{\dot\gamma}_{\rm p, p\gamma} \approx - c \, \langle \sigma_{p\gamma} f \rangle \, n'_{\rm ph} 
(\epsilon'_t) \, \epsilon'_t \, \gamma_p
\label{gpdot}
\end{equation}
where $\epsilon'_t = \epsilon' / (m_e c^2)$ and $\langle \sigma_{p\gamma} f \rangle \approx 10^{-28}$~cm$^2$ 
is the inelasticity-weighted p$\gamma$ interaction cross section. The factor $n'_{\rm ph} (\epsilon'_t) \, 
\epsilon'_t$ provides a proxy for the co-moving energy density of the target photon field, $u'_{\rm t} 
\approx m_e c^2 \, n'_{\rm ph} (\epsilon'_t) \, \epsilon'_t$. Considering that the energy lost by protons 
in p$\gamma$ interactions is shared approximately equally between photons and neutrinos, the 30~TeV -- 3~PeV 
neutrino luminosity is given by

$$
L'_{\nu} \approx \frac{1}{2} N_0 \, m_p c^2 \, \int\limits_{\gamma_1}^{\gamma_2} \gamma_p^{-\alpha_p} \, 
\left\vert {\dot\gamma}_{\rm p, p\gamma} \right\vert \, d\gamma_p 
$$
\begin{equation}
\approx 1.3 \times 10^{-14} \, N_0 \, u'_t \, {\rm cm}^3 \, {\rm s}^{-1}
\label{Lnu1}
\end{equation}
for $\alpha_p=2$.
The limits in the integral in Eq.~(\ref{Lnu1}) are given by $\gamma_1 = 6.4 \times 10^4 (D_{1} \xi_{0.05})^{-1}$
and $\gamma_2 = 6.4 \times 10^6 (D_{1} \xi_{0.05})^{-1}$, and $\alpha_p = 2$. Setting the neutrino 
luminosity from Eq.~(\ref{Lnu1}) equal to the co-moving neutrino luminosity inferred from IceCube observations, 
yields 

\begin{equation}
N_0 \, u'_t \approx 1.3 \times 10^{57} \, D_1^{-4} \, {\rm erg} \, {\rm cm}^{-3}
\label{N0_ut}
\end{equation}

We now have two independent constraints on the target photon field: First is the direct observational
constraint that the observed photon flux corresponding to the target photon field can not exceed the  observed flux in the relevant energy range. This provides a direct upper limit on $u'_t$, and will 
be used in Sect.~\ref{subsec:comoving}, \ref{subsec:stationary}. 

Second is the limit on the $\gamma\gamma$ opacity provided by the target photon field. In Sect.~
\ref{sec:Cascading}, we had found that, in order not to violate constraints from the observed
optical, X-ray and/or {\it Fermi}-LAT flux, the system needs to be either in a regime
where cascades are Compton dominated (as the minimal level of the Compton cascade emission turned out 
to lie always below the observed SED, irrespective of the value of $\tau_{\gamma\gamma, \rm max}$),
or synchrotron (or Compton-synchrotron) dominated for 
$\tau_{\gamma\gamma,max} \sim$ a few 10 (100) for small $D\leq 10$ (large $D\gg 10$) Doppler factor. 
Compton-supported cascades will occur if

\begin{equation}
{u'_t}^{\rm Comp} \gg u'_B \approx 4 \, B_1^2 \; {\rm erg \, cm}^{-3}.
\label{utCompton}
\end{equation}
in the co-moving frame with $B=B_1 10$G the magnetic field strength.

Using a simple $\delta$-function approximation for the $\gamma\gamma$ pair production cross section, we can estimate 
$\tau_{\gamma\gamma, \rm max} \sim \sigma_T \, u'_t \, R' / (3 \, m_e c^2) \sim 2.7 \times 10^{-3}
\, u_0 \, R_{16}$, where $u_0 = u'_t / ({\rm erg \, cm}^{-3})$ parameterizes the target photon energy
density, and $R_{\rm 16} = R' / (10^{16} \, {\rm cm})$ is the size of the emission region. Thus,
the condition $\tau_{\gamma\gamma,max} \sim$ a few 10 (100) for $D\leq 10$ ($D\gg 10$) translates into a limit of

$$
{u'_t}^{\rm Syn} \lesssim 10^{4} \, R_{16}^{-1} \; {\rm erg \, cm}^{-3} \hspace*{0.2cm} \mbox{for} \hspace*{0.2cm} D\leq 10,
$$
\begin{equation}
{u'_t}^{\rm Syn} \lesssim 10^{5} \, R_{16}^{-1} \; {\rm erg \, cm}^{-3} \hspace*{0.2cm} \mbox{for} \hspace*{0.2cm} D\gg 10.
\label{utSynchrotron}
\end{equation}
Combining with $u'_t \leq u'_B\approx 4 \, B_1^2 \; {\rm erg \, cm}^{-3}$ for Synchrotron-(Compton-synchrotron) 
cascades one gets:
\begin{equation}
R_{16}B_1^2 \geq 2.5\cdot 10^3 ( 2.5\cdot 10^4)  \hspace*{0.2cm} \mbox{for} \hspace*{0.2cm} D\leq 10 (D\gg 10)
\label{RBSynchrotron}
\end{equation}
This indicates the need for large field strengths and/or emission regions in the case of Synchrotron-(Compton-synchrotron) 
cascades operating in the source.

A limit on the proton kinetic power can then be derived by combining Eq.~\ref{N0_ut} with Eq.~\ref{utSynchrotron} in 
case Synchrotron-(Compton-synchrotron) cascades determine the dissipation process.
The co-moving relativistic proton energy density, assuming that the proton spectrum extends as an unbroken power-law 
with index $\alpha_p = 2$ from $\gamma_{\rm min} = 1$ 
to $\gamma_2 \gg \gamma_{\rm min}$,  is calculated by
$$
u'_p = m_p c^2 V^{'-1}N_0 \int_{1}^{\gamma_2}d\gamma_p \gamma_p^{1-\alpha_p} 
$$
\begin{equation}
\approx 3.6 \times 10^{-52} \, N_0 \, R_{16}^{-3} \, (15.7-\eta) \, {\rm erg} \, {\rm cm}^{-3}
\label{up}
\end{equation} 
with $\eta=\ln(D_1\xi_{0.05})$, and $V'$ the co-moving volume of the emission region. Since typically $\eta\approx 0$ with only small deviations expected, we will 
neglect the $\eta$-term in the following.
The relativistic proton kinetic power $L_p = 2\pi R^{'2} c \Gamma^2 u'_p$ carried by the jet in
 highly magnetized environments where pair synchrotron losses can not be neglected can then be evaluated as
\begin{equation}
L_p \gg 1.4 (14)\cdot 10^{47} (\Gamma_1/D_1^2)^2 \, {\rm erg} \, {\rm s}^{-1} \hspace*{0.2cm} \mbox{for} 
\hspace*{0.2cm} D\gg 10 (D\leq 10),
\label{LpSynchrotron}
\end{equation} 
with $\Gamma$ the bulk Lorentz factor $\Gamma \equiv 10 \, \Gamma_1$,
and a magnetic field power (using Eq.\ref{RBSynchrotron}) of
\begin{equation}
L_B \gg 0.2 (2)\cdot 10^{50} \Gamma_1^2 R_{16} \, {\rm erg} \, {\rm s}^{-1} \hspace*{0.2cm} \mbox{for} 
\hspace*{0.2cm} D\leq 10 (D\gg 10).
\label{LB}
\end{equation}
Hence, the total jet power turns out to be in this case at least of order 
$\sim 10^{49 \ldots 50}$erg\,s$^{-1}$ (or higher), independent of the origin of the target photon field. 

We now use the observed SED in the energy range of the target photons 
(here $\epsilon'_{\rm min} = 1 \ldots 170$~keV,  $\epsilon'_{\rm max} = 4\ldots 1000$~keV for D=$1\ldots 50$) to 
derive a direct limit on $u'_t$ which will then be 
combined with the previous constraints.

\subsection{Case a: Co-Moving Target Photon Field}
\label{subsec:comoving}

If the target photon field is co-moving with the emission region, the required target photon energy corresponds
to hard X-rays at an observed energy of $\epsilon_{\rm obs} \geq 16 D_{1}^2 \xi_{0.05}/E_{14}$~keV. The 
observed X-ray flux from TXS 0506+056 is of the order of $F^{\rm obs}_X \sim 10^{-12}$~erg~cm$^{-2}$~s$^{-1}$, 
implied from archival data. 
The flux received directly from the target photon field, present in an assumed spherical emission region of 
size $R'_t \equiv 10^{16} \, R_{t, 16}$~cm is $F^{\rm obs}_a = \frac{R^{'2}_t}{d_L^2} \, c \, u'_t \, D^4$ and
must be equal to or less than the observed value. This constrains the co-moving target photon energy density 
to be
\begin{equation}
u'_t \lesssim 9 \times 10^{-4} \, R_{t, 16}^{-2} \, D_1^{-4} \, {\rm erg} \, {\rm cm}^{-3}
\label{utlimita}
\end{equation}
and in turn the proton-spectrum normalization factor in Eq.~\ref{N0_ut}:
\begin{equation}
N_0 \gtrsim  1.5 \times 10^{60} \, R^{2}_{t, 16} 
\label{N0limita}
\end{equation}
The corresponding total kinetic power carried by the jet can then be evaluated as
\begin{equation} 
L_p \gtrsim 1.55 \times 10^{55} \, \Gamma_1^2 \, R_{t, 16} \; {\rm erg \; s}^{-1}.
\label{Lkina}
\end{equation}
which appears unreasonably high to be powered by an AGN accretion flow, as it exceeds the Eddington luminosity of even the most massive known supermassive black holes ($M_{\rm bh} \lesssim 10^{10} \, M_{\odot}$) 
by several orders of magnitude. 

The photon energy density limit from Eq. (\ref{utlimita}) can also be used to calculate the proton 
cooling time scale due to photopion production:
$$
t^{\rm obs}_{p\gamma} = \frac{m_e c^2}{D \, c \, \langle \sigma_{p\gamma} f \rangle \, u'_t} 
\gtrsim 3 \times 10^{13} \, R^{2}_{t, 16} \, D_1^3 \; {\rm s}
$$
\begin{equation}
\sim 9.5 \times 10^5 \, R^{2}_{t,16} D_1^3 \; {\rm yr}
\label{tpgammaa}
\end{equation}
to compare with proton synchrotron cooling. The importance of proton synchrotron radiation in 
the context of hadronic AGN models
has been first noticed by \cite{{Muecke00},{Aharonian00},{Muecke01}}. Eq.~\ref{tpgammaa} indicates 
that photopion production is several orders of magnitude less efficient than proton 
synchrotron radiation with a cooling time scale of
\begin{equation}
t^{\rm obs}_{psyn}\simeq 9 \times 10^{8} \, B^{-2}_{1} \, D_1^{-1} \left( \frac{\gamma}{6 \times 10^6} \right)\; {\rm s} \; ,
\label{tpsyna}
\end{equation}
even for protons of moderate Lorentz factors of $\gamma_p \sim 6 \times 10^6$. 
In fact, this allows us to calculate the relative radiative power output in proton synchrotron versus 
photo-pion induced emissions:
\begin{equation}
\frac{L_{\rm psy}}{L_{p\gamma}} \sim \frac{t'_{p\gamma}}{t'_{psy}} \sim 4 \times 10^4 \, R^{2}_{t, 16} \, 
D_1^4 \, B_1^2 \left( \frac{\gamma}{6 \times 10^6} \right)
\label{Lratioa}
\end{equation}
with proton synchrotron emission peaking at

\begin{equation}
\nu_{\rm psy}^{\rm obs} \sim 9 \times 10^{16} \, B_1 \, \left( \frac{\gamma}{6 \times 10^6} \right)^2 \, 
D_1 \, {\rm Hz}
\label{nupsya}
\end{equation}
i.e., in the EUV/soft X-ray band.
This implies that, in this scenario, the protons producing the 0.03-3~PeV neutrino flux are expected to 
produce proton synchrotron emission in the X-ray regime at a $\nu F_{\nu}$ flux value larger than the 
corresponding $\nu F_{\nu}$ neutrino flux value by a factor given by Eq. (\ref{Lratioa}). Reducing 
this to the limit based on the observed X-ray flux ($L_{\rm psy} / L_{p\gamma} \lesssim 0.1$) would 
require an unusually low magnetic field value ($B \lesssim 1$~G) and/or a very low Doppler factor 
$D \sim 1$. This is demonstrated in Figs.~6, 13 where we have added the proton synchrotron radiation
on top of the minimal cascade component for various model parameters.

Based on these results, we may confidently rule out a scenario in which the IceCube neutrino flare 
flux from TXS~0506+056 is photo-hadronically produced in a target photon field that is co-moving with
the emission (and neutrino production) region.

\subsection{Case b: Stationary Target Photon Field in the AGN Frame}
\label{subsec:stationary}

In the case of a target photon field that is stationary in the AGN rest frame, the target photon 
energy (in the AGN frame) should be $\epsilon_{\rm obs} \geq 0.16 D_1 \xi_{0.05}/E_{14}$~keV, and the 
AGN-frame radiation energy density is boosted into the co-moving frame as $u'_t \approx \Gamma^2 \, u_t$. 
In the external-photon-field case, 
the spatial extent of the photon field may be scaled in units of a typical BLR size, $R_t \equiv 10^{17} 
\, R_{t, 17}$~cm (AGN frame), and the corresponding flux received is $F_{UV}^{\rm obs} = 
\frac{R_t^2}{d_L^2} \, c \, u'_t \, \Gamma^{-2}$.

For the Compton-dominated cascade case we now insert the energy density from Eq.~\ref{utCompton} to derive
\begin{equation}
F_{UV}^{\rm obs} \gg 4.3\cdot 10^{-13} (B_1 R_{t,17} / \Gamma_1)^2 {\rm erg} \, {\rm cm}^{-2} \, {\rm s}^{-1}
\label{Compdomb}
\end{equation}
which is below the observed SED from archival data for not too large field strengths.
Inspecting now the case for a highly magnetized environment where synchrotron losses become non-negliglible we derive
a UV -- soft X-ray flux of
\begin{equation}
F_{UV}^{\rm obs} \gg (10^{-9}\ldots 10^{-8}) R_{16}^{-1} (R_{t,17} / \Gamma_1)^2 {\rm erg} \, {\rm cm}^{-2} \, {\rm s}^{-1}
\label{Syndomb}
\end{equation}
for $D\leq 10$ ($D\gg 10$) by making use of Eq.~\ref{utSynchrotron}. This flux level seems several orders of magnitude 
higher than the observed SED from archival data implies, and hence gives strong arguments to rule out an environment of
the neutrino emission region where Synchrotron-(Compton-synchrotron) cascades operate.

From the observed SED a conservative upper limit on the $\nu F_\nu$ flux at UV -- soft X-rays may be placed at 
$\nu F_\nu\lesssim 10^{-11}$ erg cm$^{-2}$ s$^{-1}$. 
The co-moving target photon energy density can then be constrained as

\begin{equation}
u'_t \lesssim 94 \, \Gamma_1^2 \, R_{t, 17}^{-2} \; {\rm erg} \, {\rm cm}^{-3}
\label{utlimitb}
\end{equation}
giving a corresponding pair production optical depth of
\begin{equation}
\tau_{\gamma\gamma,\rm max} \lesssim 0.25 \, R_{16} \, \Gamma_1^2 \, R_{t, 17}^{-2}
\label{taugglimitb}
\end{equation}
and a photohadronic optical depth of
\begin{equation}
\tau_{p\gamma,\rm max} \lesssim 8\cdot 10^{-4} \, R_{16} \, \Gamma_1^2 \, R_{t, 17}^{-2}
\label{taupglimitb}
\end{equation}
above the thresholds of the interactions.
For moderate magnetic fields, $B \lesssim 10$~G, this may lead to Compton-supported cascades, thus
not violating limits on the multi-wavelength cascade flux. Eq. (\ref{N0_ut}) then constrains the 
proton-spectrum normalization to

\begin{equation}
N_0 \gtrsim 1.4 \times 10^{55} \, D_1^{-4} \, \Gamma_1^{-2} \, R_{t, 17}^2
\label{N0limitb}
\end{equation}
yielding a required proton kinetic jet power of

\begin{equation}
L_{p} \gtrsim 1.5 \times 10^{50} \, D_1^{-4} \, R_{t, 17}^2 \, R_{16}^{-1}
\; {\rm erg \; s}^{-1}. 
\label{Lkinb}
\end{equation}
Due to the strong inverse dependence on the Doppler factor, for $D$ significantly exceeding $10$ (and/or 
a larger target-photon-field radius or smaller emission-region size), this may be within the range 
plausibly powered by accretion onto a supermassive black hole of mass $\sim 3\cdot 10^8$M$_\sun$ \citep{Padovani19}. The relativistic proton power from
Eq. (\ref{Lkinb}) may be compared to the power carried along the jet in magnetic fields, 

\begin{equation}
L_B = 3.8 \times 10^{45} \, R_{16}^2 \, \Gamma_1^2 \, B_1^2 \; {\rm erg \, s}^{-1}.
\label{LBb}
\end{equation}
Thus, for $B \lesssim 10$~G, the jet would have to be strongly dominated by kinetic power of the
relativistc protons. 

For the target photon field of Eq. (\ref{utlimitb}), the proton photopion cooling time scale is, analogous to
Eq. (\ref{tpgammaa}),

\begin{equation}
t^{\rm obs}_{p\gamma} \gtrsim 2.9 \times 10^9 \, \Gamma_1^{-2} \, R_{17}^2 \; {\rm s} \sim 92 \, 
\Gamma_1^{-2} \, R_{17}^2 \; {\rm yr}. 
\label{tpgammab}
\end{equation}

Also in this case, proton-synchrotron radiation, at X-ray frequencies according to
Eq. (\ref{nupsya}), can not be neglected, and the ratio of proton-synchrotron to photo-pion 
induced radiative (and neutrino) output will be

\begin{equation}
\frac{L_{\rm psy}}{L_{p\gamma}} \gtrsim 0.4 \, \Gamma_1^{-2} \, D_1 \, R_{t,17}^2 \, B_1^2 \, 
\left( \frac{\gamma}{6 \times 10^6} \right)
\label{Lratiob}
\end{equation}
thus predicting X-ray proton-synchrotron emission with a flux of $\nu F_{\nu} (X) 
\gtrsim 10^{-10} \, \Gamma_1^{-2} \, D_1 \, R_{t,17}^2 \, B_1^2 \, 
(\gamma / [6 \times 10^6])$~erg~cm$^{-2}$~s$^{-1}$. For a magnetic field of $B \sim$ a few G, 
this does not seem to violate observational limits (see also Fig.~6). 

We thus conclude that a scenario in which an external soft X-ray photon field provides the targets
for photo-pion reactions producing the IceCube neutrinos of TXS~0506+056 in an environment where 
dominantly Compton scattering supports the pair cascading, appears feasible. Eq.~\ref{taupglimitb} 
and  Eq.~\ref{taugglimitb} imply 
in this case still very inefficient neutrino production, and a corresponding pair cascade flux that 
lies significantly below the 
gamma-ray flux level observed by the LAT during the neutrino flare (see Figs.~\ref{fig:4} -- \ref{fig:8}). 
Further high energy radiation processes are therefore needed to explain the observed gamma-ray 
SED, which is unlikely proton-initiated, during the neutrino flare.

\begin{table*}[h!]
\centering
\begin{tabular}{c|c|l}
\hline
Origin of target photon field & Further setup constraints & Results\\
\hline
All target photon fields &  $u'_t\gg u'_B$  &  no constraints  \\
(cascading constraints) &  $u'_t\leq u'_B$ &  $\tau_{\gamma\gamma,\rm{max}}\sim$ (a few) $10$ for $D\leq 10$    \\
                        &                  &  $\tau_{\gamma\gamma,\rm{max}}\sim$ (a few) $10^{1...2}$ for $D\gg 10$    \\
\hline
Co-moving target photon field &          &  {\it ruled out:} too high $L_{p}$ (Eq.~\ref{Lkina});\\
                              &          &  $p$ syn overshoots X-ray archival flux (Eq.~\ref{Lratioa},\ref{nupsya})   \\
\hline
Stationary target photon field & $u'_t\leq u'_B$ & {\it ruled out:} target photon flux\tablenotemark{a} overshoots UV/ \\
                               &                 & soft X-ray archival flux (Eq.~\ref{Syndomb})\\
                               & $u'_t\gg u'_B$  & {\it viable:} target photon flux below UV/\\
                               &                & soft X-ray archival flux (Eq.~\ref{Compdomb});\\
                               &                & $L_{p}$ (Eq.~\ref{Lkinb}) acceptable for sufficiently large $D$\\  
\hline
\end{tabular}
\tablenotetext{a}{Includes cascading constraints.}
\caption{Summary of tested setups and results.} 
\label{tab:summary}
\end{table*}

\section{Discussion and Conclusions}
\label{sec:summary}

In this work we present a procedure to derive constraints on the broadband photon spectral energy
distribution for a given (observed) neutrino spectrum and in the framework of photo-hadronically 
produced neutrinos from relativistic protons accelerated in jetted AGN. This procedure can accommodate any 
origin for the dominant target photon field. For this purpose we first reconstruct, in the 
co-moving jet-frame, the minimum target photon spectrum required to produce the neutrino spectrum, and
calculate all corresponding further secondary particles produced through these interactions (photomeson 
and Bethe-Heitler pair production). A rather narrow energy range at hard X-rays (in the co-moving jet frame) 
is sufficient to explain the 2014/15 IceCube neutrino flare spectrum from the 
direction of TXS~0506+056 by photomeson production of relativistic protons from an injection spectrum with index 
$\alpha_p=2.0\pm^{0.1}_{0.2}$ extending to PeV -- hundreds of PeV energies. For typical spatial scales of the emission region 
and magnetic field strengths these protons are well contained in this region. 

The produced high-energy photons and electron-positron pairs initiate 
electromagnetic cascades in the neutrino emission region with an efficiency that is directly linked to
the neutrino production rate. Both the energy-dependent pair production optical depth and the 
radiation process (inverse Compton scattering, synchrotron radiation) that feed the cascades 
determine the resulting shape of the photon spectrum escaping the source. The so derived photon 
spectrum, for a given Doppler factor, can be considered as the minimum radiation emerging from 
the source that is strictly associated with the photo-hadronically produced neutrinos. 
By comparing our simulated cascade spectra to broadband SEDs from 
observations carried out quasi-simultaneous to the neutrino detection, one can then derive
constraints on the environment of the neutrino-producing site. A summary of the tested setups and resulting constraints is provided in Tab.~1.

A comprehensive study of 
these (steady-state) minimum cascade SEDs, when operating in a narrow X-ray target photon field, 
reveals rather low flux levels at GeV-energies if inverse Compton radiation feeds the cascade, even
for an optically thin radiative environment. In case of highly magnetized environments where 
pair synchrotron losses can not be neglected, the synchrotron/Compton-synchrotron 
supported cascades extend across a much wider energy range than the Compton-supported cascades, down to radio/optical frequencies. 
This allows deep optical to X-ray measurements to set limits on the efficiency of neutrino production in this region.

The efficient photo-pion production of $\sim$~PeV neutrinos
requires a high target-photon density, leading to $\gamma\gamma$ absorption depths $\tau_{\gamma\gamma}
\gg 1$ for $\sim$~GeV photons. We therefore concluded that efficient photo-pion production of IceCube neutrinos in 
blazars does not predict a causal connection between quasi-contemporaneous GeV $\gamma$-ray and neutrino 
emission here. 

We then combined the derived limits on the $\gamma\gamma$ opacity from the cascade SEDs and co-moving 
neutrino luminosity of TXS~0506+056 with direct observational constraints from its MWL SED in the expected 
energy range of the required minimum target photon field. Considering first the extreme case of a co-spatially 
produced (co-moving) target photon field
for photo-pion production of the neutrino flare of TXS~0506+056, we deduced extreme values for the required 
proton kinetic power and a too high proton synchrotron component that would violate the observed photon SED 
at X-ray energies. This makes such a scenario implausible. Also regarding the case of a stationary 
(in the AGN frame) target photon field as the other extreme case results in too high expected flux values in 
the soft X-ray regime as compared to the observed SED if synchrotron or Compton-synchrotron cascades operate 
in the neutrino emission region.

The only viable scenario appears to be a stationary soft X-ray photon field providing the targets for 
photo-pion production of the observed neutrinos with the resulting cascades being fed by inverse Compton 
radiation only. The required proton kinetic 
powers appear in a plausible range for a jet powered by accretion onto a supermassive black hole, and the 
expected proton synchrotron component
does not exceed the observed X-ray flux for not too large field strengths. 
In this environment, however, neutrino production is very inefficient ($\tau_{p\gamma}\sim 10^{-3}$ for 
typical blazar model parameters) during the neutrino flare
and the associated pair cascade GeV-flux too low to explain the LAT observations. 
The origin of the LAT spectrum observed during the neutrino flare period can therefore not be associated 
with the neutrino producing mechanism, and the GeV-flux has to be emitted from zones in the jet other than the neutrino zone.
Any link between the neutrino and GeV-flux is then determined by a corresponding connection between the $\gamma$-ray and
neutrino emitting zones.

\acknowledgements

The work of M.B. is supported through the South African Research Chair Initiative of the National 
Research Foundation\footnote{Any opinion, finding and conclusion or recommendation expressed in 
this material is that of the authors and the NRF does not accept any liability in this regard.} 
and the Department of Science and Technology of South Africa, under SARChI Chair grant No. 64789. A.R. acknowledges financial support from the Austrian Science Fund (FWF), project I 3452-N27. 
 
SB was supported by the NASA Postdoctoral Program.

We are grateful to Valentina La Parola, Alberto Segreto and Marco Ajello for being 
instrumental in deriving the {\it BAT} upper limit used in this work. We also thank Matthew Baring for 
providing valuable comments that lead to significant improvements of the manuscript, and P. Padovani and A. Tohuvavohu for valuable discussions on the manuscript text. We also thank K. Murase who pointed out to us the work Murase et al (2018), and S. Gao who pointed out to us the work Gao et al (2017).

We thank the ASAS-SN team for providing publicly, real time optical data used in this work.

The \textit{Fermi}-LAT Collaboration acknowledges generous ongoing support
from a number of agencies and institutes that have supported both the
development and the operation of the LAT as well as scientific data analysis.
These include the National Aeronautics and Space Administration and the
Department of Energy in the United States, the Commissariat \`a l'Energie Atomique
and the Centre National de la Recherche Scientifique / Institut National de Physique
Nucl\'eaire et de Physique des Particules in France, the Agenzia Spaziale Italiana
and the Istituto Nazionale di Fisica Nucleare in Italy, the Ministry of Education,
Culture, Sports, Science and Technology (MEXT), High Energy Accelerator Research
Organization (KEK) and Japan Aerospace Exploration Agency (JAXA) in Japan, and
the K.~A.~Wallenberg Foundation, the Swedish Research Council and the
Swedish National Space Board in Sweden.

Additional support for science analysis during the operations phase is gratefully
acknowledged from the Istituto Nazionale di Astrofisica in Italy and the Centre
National d'\'Etudes Spatiales in France. This work performed in part under DOE
Contract DE-AC02-76SF00515.

\software{Fermipy \citep[v0.17.3, ][]{wood17}, batimager \citep{segreto10}, SOPHIA \citep{SOPHIA}, \citet{Protheroe96}}



\end{document}